\documentclass[aps,prd,final,nofootinbib,twocolumn,showpacs]{revtex4-1}

\usepackage{amssymb}
\usepackage{graphicx}                            
\usepackage{amsmath} 
\usepackage{color}
\usepackage{perpage} 

\newcommand{\abs}[1]{\lvert#1\rvert}

\newcommand{\beq}{\begin{equation}}
\newcommand{\beqn}{\begin{eqnarray}}
\newcommand{\eeq}{\end{equation}}
\newcommand{\eeqn}{\end{eqnarray}}

\begin{document}
\title{Multifield Inflation after \emph{Planck}: Isocurvature Modes from Nonminimal Couplings }
\author{Katelin~Schutz, Evangelos~I.~Sfakianakis and David~I.~Kaiser}
\email{kschutz14@mit.edu, esfaki@mit.edu, dikaiser@mit.edu}
\date{\today} 
\affiliation{Center for Theoretical Physics and Department of Physics, \\
Massachusetts Institute of Technology, Cambridge, Massachusetts 02139 USA}
\begin{abstract}
\noindent
Recent measurements by the {\it Planck} experiment of the power spectrum of temperature anisotropies in the cosmic microwave background radiation (CMB) reveal a deficit of power in low multipoles compared to the predictions from best-fit $\Lambda$CDM cosmology. If the low-$\ell$ anomaly persists after additional observations and analysis, it might be explained by the presence of primordial isocurvature perturbations in addition to the usual adiabatic spectrum, and hence may provide the first robust evidence that early-universe inflation involved more than one scalar field. In this paper we explore the production of isocurvature perturbations in nonminimally coupled two-field inflation. We find that this class of models readily produces enough power in the isocurvature modes to account for the \emph{Planck} low-$\ell$ anomaly, while also providing excellent agreement with the other {\it Planck} results.
\end{abstract}
\pacs{04.62+v; 98.80.Cq. Published in {\it Physical Review D} {\bf 89}: 064044 (2014)}
\maketitle
\section{Introduction}

Inflation is a leading cosmological paradigm for the early universe, consistent with the myriad of observable quantities that have been measured in the era of precision cosmology \cite{InflationArticles,InflationReviews1,InflationMPertsReviews}. However, a persistent challenge has been to reconcile successful inflationary scenarios with well-motivated models of high-energy physics. Realistic models of high-energy physics, such as those inspired by supersymmetry or string theory, routinely include multiple scalar fields at high energies \cite{InflationHEPReviews}. Generically, each scalar field should include a nonminimal coupling to the spacetime Ricci curvature scalar, since nonminimal couplings arise as renormalization counterterms when quantizing scalar fields in curved spacetime \cite{NMC,BirrellDavies,Buchbinder,Faraoni2000}. The nonminimal couplings typically increase with energy-scale under renormalization-group flow \cite{Buchbinder}, and hence should be large at the energy-scales of interest for inflation. We therefore study a class of inflationary models that includes multiple scalar fields with large nonminimal couplings.

It is well known that the predicted perturbation spectra from single-field models with nonminimal couplings produce a close fit to observations. Following conformal transformation to the Einstein frame, in which the gravitational portion of the action assumes canonical Einstein-Hilbert form, the effective potential for the scalar field is stretched by the conformal factor to be concave rather than convex \cite{NMCsingle,HiggsInflation}, precisely the form of inflationary potential most favored by the latest results from the {\it Planck} experiment \cite{PlanckInflation}. 

The most pronounced difference between multifield inflation and single-field inflation is the presence of more than one type of primordial quantum fluctuation that can evolve and grow. The added degrees of freedom may lead to observable departures from the predictions of single-field models, including the production and amplification of isocurvature modes during inflation \cite{IsocurvMultifieldOriginal,BassettWandsTurn,WandsBartolo,Mazumdar,WandsReview,TurnCovariant,PTGeometric,KS}.

Unlike adiabatic perturbations, which are fluctuations in the energy density, isocurvature perturbations arise from spatially varying fluctuations in the local equation of state, or from relative velocities between various species of matter. When isocurvature modes are produced primordially and stretched beyond the Hubble radius, causality prevents the redistribution of energy density on super-horizon scales. When the perturbations later cross back within the Hubble radius, isocurvature modes create pressure gradients that can push energy density around, sourcing curvature perturbations that contribute to large-scale anisotropies in the cosmic microwave background radiation (CMB). (See, e.g., \cite{CMBIsocurvature,PlanckInflation}.) 

The recent measurements of CMB anisotropies by {\it Planck} favor a combination of adiabatic and isocurvature perturbations in order to improve the fit at low multipoles ($\ell \sim 20 - 40$) compared to the predictions from the simple, best-fit $\Lambda$CDM model in which primordial perturbations are exclusively adiabatic. The best fit to the present data arises from models with a modest contribution from isocurvature modes, whose primordial power spectrum ${\cal P}_{\cal S} (k)$ is either scale-invariant or slightly blue-tilted, while the dominant adiabatic contribution, ${\cal P}_{\cal R} (k)$, is slightly red-tilted \cite{PlanckInflation}. The low-$\ell$ anomaly thus might provide the first robust empirical evidence that early-universe inflation involved more than one scalar field.

Well-known multifield models that produce isocurvature perturbations, such as axion and curvaton models, are constrained by the {\it Planck} results and do not improve the fit compared to the purely adiabatic $\Lambda$CDM model \cite{PlanckInflation}. As we demonstrate here, on the other hand, the general class of multifield models with nonminimal couplings can readily produce isocurvature perturbations of the sort that could account for the low-$\ell$ anomaly in the {\it Planck} data, while also producing excellent agreement with the other spectral observables  measured or constrained by the {\it Planck} results, such as the spectral index $n_s$, the tensor-to-scalar ratio $r$, the running of the spectral index $\alpha$, and the amplitude of primordial non-Gaussianity $f_{\rm NL}$.

Nonminimal couplings in multifield models induce a curved field-space manifold in the Einstein frame \cite{DKconf}, and hence one must employ a covariant formalism for this class of models. Here we make use of the covariant formalism developed in \cite{KMS}, which builds on pioneering work in \cite{BassettWandsTurn,PTGeometric}. In Section \ref{mod} we review the most relevant features of our class of models, including the formal machinery required to study the evolution of primordial isocurvature perturbations. In Section \ref{traj} we focus on a regime of parameter space that is promising in the light of the {\it Planck} data, and for which analytic approxmations are both tractable and in close agreement with numerical simulations. In Section \ref{results} we compare the predictions from this class of models to the recent {\it Planck} findings. Concluding remarks follow in Section \ref{conclusions}.

\section{Model}
\label{mod}
We consider two nonminimally coupled scalar fields $\phi^I\, \epsilon\, \{\phi, \, \chi\}$. We work in 3+1 spacetime dimensions with the spacetime metric signature $(-, \, +, \, +, \, +)$. We express our results in terms of the reduced Planck mass, $M_{\text{\rm pl}} \equiv \left( 8 \pi G \right) ^{-1/2} = $ 2.43 $\times$ 10$^{18}$ GeV. Greek letters ($\mu$, $\nu$) denote spacetime 4-vector indices, lower-case Roman letters ($i$, $j$) denote spacetime 3-vector indices, and capital Roman letters ($I$, $J$) denote field-space indices. We indicate Jordan-frame quantities with a tilde, while Einstein-frame quantities will be \emph{sans} tilde. Subscripted commas indicate ordinary partial derivatives and subscripted semicolons denote covariant derivatives with respect to the spacetime coordinates.

We begin with the action in the Jordan frame, in which the fields' nonminimal couplings remain explicit: 
\begin{equation}
\tilde{S} = \int d^4 x \sqrt{-\tilde{g}} \left[  f(\phi^I) \tilde{R} - \frac{1}{2} \tilde{\mathcal{G}}_{I J} \tilde{g}^{\mu \nu} \partial_{\mu} \phi^I \partial_{\nu} \phi^J - \tilde{V}(\phi^I) \right],
\end{equation}
where $\tilde{R}$ is the spacetime Ricci scalar, $f(\phi^I)$ is the nonminimal coupling function, and $\tilde{\mathcal{G}}_{I J}$ is the Jordan-frame field space metric. We set $\tilde{\mathcal{G}}_{I J} = \delta_{IJ}$, which gives canonical kinetic terms in the Jordan frame. We take the Jordan-frame potential, $\tilde{V}(\phi^I)$, to have a generic, renormalizable polynomial form with an interaction term:
 \begin{equation}
\tilde{V}(\phi, \chi) = \frac{\lambda_{\phi}}{4} \phi^4 + \frac{g}{2} \phi^2 \chi^2 + \frac{\lambda_{\chi}}{4} \chi^4,
\end{equation}
with dimensionless coupling constants $\lambda_I$ and $g$. As discussed in \cite{KMS}, the inflationary dynamics in this class of models are relatively insensitive to the presence of mass terms, $m_\phi^2 \phi^2$ or $m_\chi^2 \chi^2$, for realistic values of the masses that satisfy $m_\phi , m_\chi \ll M_{\rm pl}$. Hence we will neglect such terms here.

\subsection{Einstein-Frame Potential}
\label{ff}

We perform a conformal transformation to the Einstein frame by rescaling the spacetime metric tensor, 
\begin{equation}
\tilde{g}_{\mu \nu}(x) = \Omega^2(x)\, g_{\mu \nu} (x) ,
\end{equation}
where the conformal factor $\Omega^2(x)$ is related to the nonminimal coupling function via the relation
\begin{equation}
\Omega^2(x) = \frac{2}{M_{\text{\rm pl}}^2} f\left(\phi^I (x)\right).
\end{equation}
This transformation yields the action in the Einstein frame,
\begin{equation}
S = \int d^4 x \sqrt{-g} \left[ \frac{M_{\text{\rm pl}}^2}{2} R - \frac{1}{2} \mathcal{G}_{I J} g^{\mu \nu} \partial_{\mu} \phi^I \partial_{\nu} \phi^J - V(\phi^I) \right],
\end{equation}
where all the terms \emph{sans} tilde are stretched by the conformal factor. For instance, the conformal transformation to the Einstein frame induces a nontrivial field-space metric \cite{DKconf}
\begin{equation}
\label{metric}
\mathcal{G}_{I J} = \frac{M_{\text{\rm pl}}^2}{2 f} \left[\delta_{IJ} + \frac{\,3\,}{f}\, f_{,I} f_{,J}\right] ,
\end{equation}
and the potential is also stretched so that it becomes
\begin{equation}
\begin{split}
V(\phi, \chi) &= \frac{M_{\rm pl}^4}{(2f)^2 } \tilde{V} (\phi, \chi) \\
&= \frac{M_{\text{\rm pl}}^4}{(2f)^2} \left[\frac{\lambda_{\phi}}{4} \phi^4 + \frac{g}{2} \phi^2 \chi^2 + \frac{\lambda_{\chi}}{4} \chi^4\right].
\end{split}
\label{VEpot}
\end{equation}
The form of the nonminimal coupling function is set by the requirements of renormalization \cite{NMC,BirrellDavies},
\begin{equation}
f(\phi, \chi) = \frac{1}{2} [M^2 + \xi_{\phi} \phi^2 +\xi_{\chi} \chi^2] ,
\label{nmc}
\end{equation}
where $\xi_\phi$ and $\xi_\chi$ are dimensionless couplings and $M$ is some mass scale such that when the fields settle into their vacuum expectation values, $f \rightarrow M_{\rm pl}^2 / 2$. Here we assume that any nonzero vacuum expectation values for $\phi$ and $\chi$ are much smaller than the Planck scale, and hence we may take $M = M_{\rm pl}$.
\begin{figure}
\centering
\includegraphics[width=3.2in]{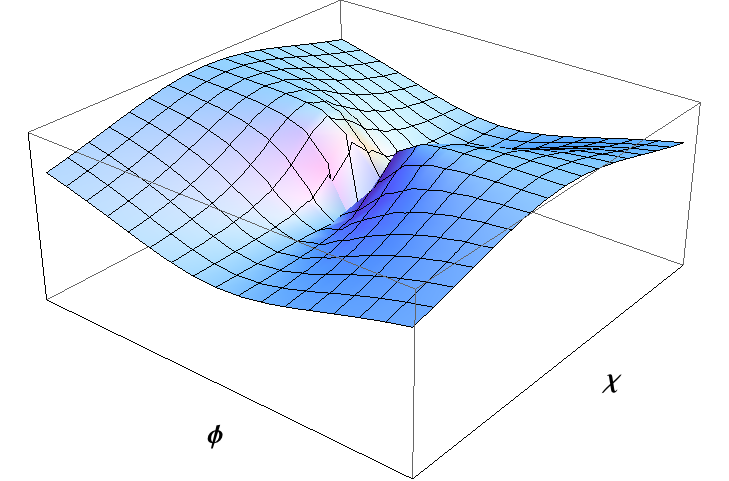} 
\caption{ Potential in the Einstein frame, $V (\phi^I )$ in Eq. (\ref{VEpot}). The parameters shown here are $\lambda_\chi = 0.75 \> \lambda_\phi$, $g = \lambda_\phi$, $\xi_\chi = 1.2 \> \xi_\phi$, with $\xi_\phi \gg 1$ and $\lambda_\phi >0$.}
\label{fig:VE}
\end{figure}

The conformal stretching of the potential in the Einstein frame makes it concave and asymptotically flat along either direction in field space, $I = \phi, \chi$,
\beq
V (\phi^I ) \rightarrow \frac{M_{\rm pl}^4}{4} \frac{\lambda_I}{ \xi_I^2} \left[ 1 + {\cal O} \left( \frac{M_{\rm pl}^2}{\xi_J (\phi^I )^2 } \right) \right] 
\label{VEasymptotic}
\eeq
(no sum on $I$). For non-symmetric couplings, in which $\lambda_\phi \neq \lambda_\chi$ and/or $\xi_\phi \neq \xi_\chi$, the potential in the Einstein frame will develop ridges and valleys, as shown in Fig. \ref{fig:VE}. Crucially, $V > 0$ even in the valleys (for $g > - \sqrt{ \lambda_\phi \lambda_\chi}$), and hence the system will inflate (albeit at varying rates) whether the fields ride along a ridge or roll within a valley, until the fields reach the global minimum of the potential at $\phi = \chi = 0$.

Across a wide range of couplings and initial conditions, the models in this class obey a single-field attractor \cite{KS}. If the fields happen to begin evolving along the top of a ridge, they will eventually fall into a neighboring valley. Motion in field space transverse to the valley will quickly damp away (thanks to Hubble drag), and the fields' evolution will include almost no further turning in field space. Within that single-field attractor, predictions for $n_s$, $r$, $\alpha$, and $f_{\rm NL}$ all fall squarely within the most-favored regions of the latest {\it Planck} measurements \cite{KS}. 

\begin{figure}
\includegraphics[width = 0.45\textwidth]{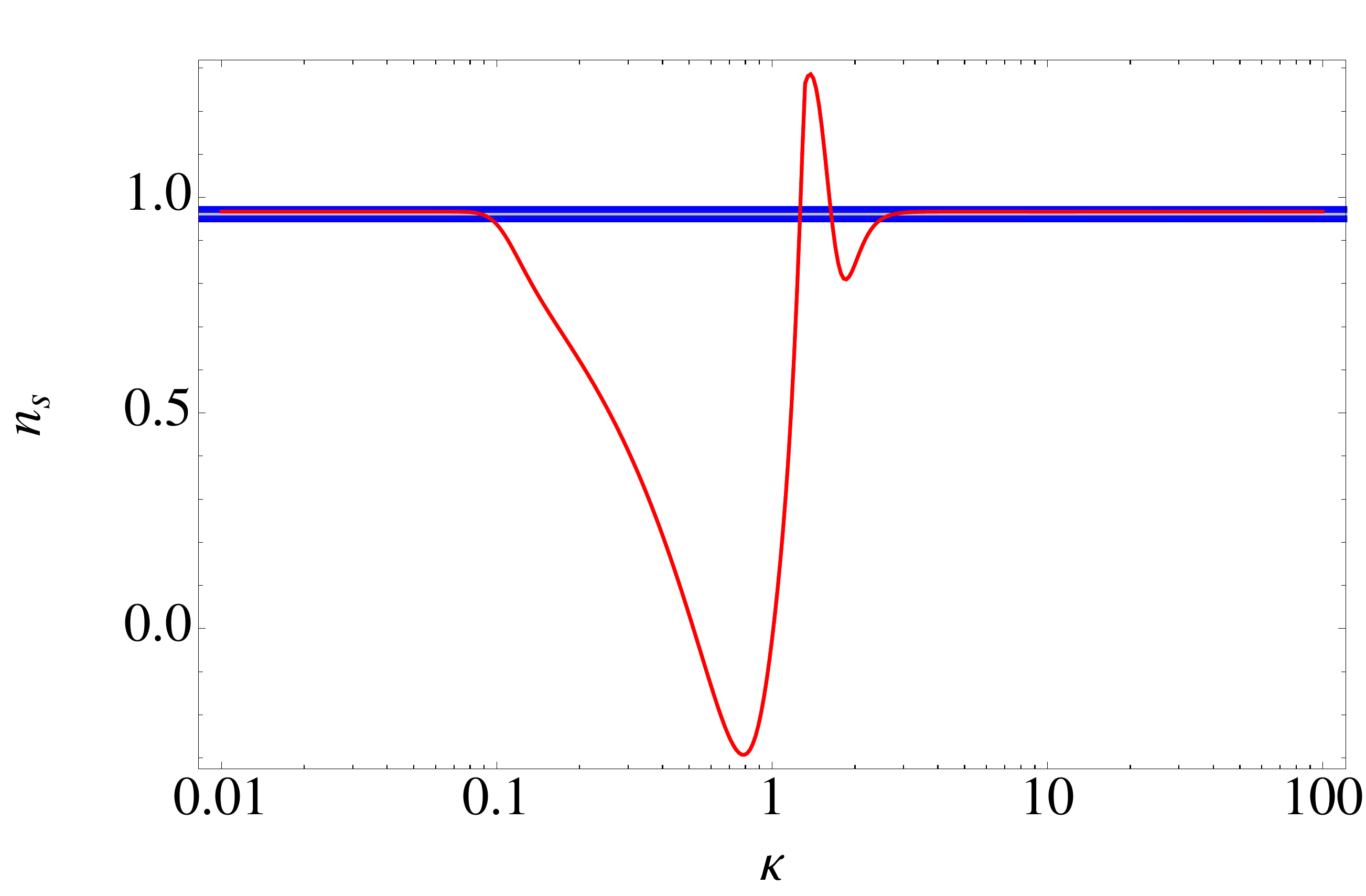}
\caption{The spectral index $n_s$ (red), as given in Eq. (\ref{nsend}), for different values of $\kappa$, which characterizes the local curvature of the potential near the top of a ridge. Also shown are the $1\sigma$ (thin, light blue) and $2\sigma$ (thick, dark blue) bounds on $n_s$ from the {\it Planck} measurements. The couplings shown here correspond to $\xi_\phi = \xi_\chi = 10^3$, $\lambda_\phi = 10^{-2}$, and $\lambda_\chi = g$, fixed for a given value of $\kappa$ from Eq. (\ref{kappa}). The fields' initial conditions are $\phi = 0.3$, $\dot{\phi}_0 = 0$, $\chi_0 = 10^{-3}$, $\dot{\chi}_0 = 0$, in units of $M_{\rm pl}$.   }
\label{fig:nswide}
\end{figure}

The fields' approach to the attractor behavior --- essentially, how quickly the fields roll off a ridge and into a valley --- depends on the local curvature of the potential near the top of a ridge. Consider, for example, the case in which the direction $\chi = 0$ corresponds to a ridge. To first order, the curvature of the potential in the vicinity of $\chi = 0$ is proportional to $(g \xi_\phi - \lambda_\phi \xi_\chi )$ \cite{KMS}. As we develop in detail below, a convenient combination with which to characterize the local curvature near the top of such a ridge is
\beq
\kappa \equiv \frac{4 (\lambda_\phi \xi_\chi - g \xi_\phi) }{\lambda_\phi } .
\label{kappa}
\eeq
As shown in Fig. \ref{fig:nswide}, models in this class produce excellent agreement with the latest measurements of $n_s$ from {\it Planck} across a wide range of parameters, where $n_s \equiv 1 + d \ln {\cal P}_{\cal R} / d \ln k$. Strong curvature near the top of the ridge corresponds to $\kappa \gg 1$: in that regime, the fields quickly roll off the ridge, settle into a valley of the potential, and evolve along the single-field attractor for the duration of inflation, as analyzed in \cite{KS}. More complicated field dynamics occur for intermediate values, $0.1 < \kappa < 4$, for which multifield dynamics pull $n_s$ far out of agreement with empirical observations. The models again produce excellent agreement with the {\it Planck} measurements of $n_s$ in the regime of weak curvature, $0 \leq \kappa \leq 0.1$.

As we develop below, other observables of interest, such as $r$, $\alpha$, and $f_{\rm NL}$, likewise show excellent fit with the latest observations. In addition, the regime of weak curvature, $\kappa \ll 1$, is particularly promising for producing primordial isocurvature perturbations with characteristics that could explain the low-$\ell$ anomaly in the recent {\it Planck} measurements. Hence for the remainder of this paper we focus on the regime $\kappa \ll 1$, a region that is amenable to analytic as well as numerical analysis.

\subsection{Coupling Constants}

The dynamics of this class of models depend upon combinations of dimensionless coupling constants like $\kappa$ defined in Eq. (\ref{kappa}) and others that we introduce below. The phenomena analyzed here would therefore hold for various values of $\lambda_I$ and $\xi_I$, such that combinations like $\kappa$ were unchanged. Nonetheless, it is helpful to consider reasonable ranges for the couplings on their own.

The present upper bound on the tensor-to-scalar ratio, $r < 0.12$, constrains the energy-scale during inflation to satisfy $H (t_{\rm hc}) / M_{\rm pl} \leq 3.7 \times 10^{-5}$ \cite{PlanckInflation}, where $H (t_{\rm hc})$ is the Hubble parameter at the time during inflation when observationally relevant perturbations first crossed outside the Hubble radius. During inflation the dominant contribution to $H$ will come from the value of the potential along the direction in which the fields slowly evolve. Thus we may use the results from {\it Planck} and Eq. (\ref{VEasymptotic}) to set a basic scale for the ratios of couplings, $\lambda_I / \xi_I^2$. For example, if the fields evolve predominantly along the direction $\chi \sim 0$, then during slow roll the Hubble parameter will be
\beq
H \simeq \sqrt{\frac{ \lambda_\phi }{12 \xi_\phi^2 } } \> M_{\rm pl} ,
\label{Hamplitude}
\eeq
and hence the constraint from {\it Planck} requires $\lambda_\phi / \xi_\phi^2 \leq 1.6 \times 10^{-8}$. 

We adopt a scale for the self-couplings $\lambda_I$ by considering a particularly elegant member of this class of models. In Higgs inflation \cite{HiggsInflation}, the self-coupling $\lambda_\phi$ is fixed by measurements of the Higgs mass near the electroweak symmetry-breaking scale, $\lambda_\phi \simeq 0.1$, corresponding to $m_H \simeq 125$ GeV \cite{HiggsDiscovery,HiggsMultifield}. Under renormalization-group flow, $\lambda_\phi$ will fall to the range $0 < \lambda_\phi < 0.01$ at the inflationary energy scale \cite{HiggsRunning}. Eq. (\ref{Hamplitude}) with $\lambda = 0.01$ requires $\xi_\phi \geq 780$ at inflationary energy scales to give the correct amplitude of density perturbations. For our general class of models, we therefore consider couplings at the inflationary energy scale of order $\lambda_I, g \sim {\cal O} (10^{-2})$ and $\xi_I \sim {\cal O} (10^3)$. Taking into account the running of both $\lambda_I$ and $\xi_I$ under renormalization-group flow, these values correspond to $\lambda_I \sim {\cal O} (10^{-1})$ and $\xi_I \sim {\cal O} (10^2)$ at low energies \cite{HiggsRunning}. 

We consider these to be reasonable ranges for the couplings. Though one might prefer dimensionless coupling constants to be ${\cal O} (1)$ in any ``natural" scenario, the ranges chosen here correspond to low-energy couplings that are no more fine-tuned than the fine-structure constant, $\alpha_{\rm EM} \simeq 1/137$. Indeed, our choices are relatively conservative. For the case of Higgs inflation, the running of $\lambda_\phi$ is particularly sensitive to the mass of the top quark. Assuming a value for $m_{\rm top}$ at the low end of the present $2\sigma$ bound yields $\lambda_\phi \simeq 10^{-4}$ rather than $10^{-2}$ at high energies, which in turn requires $\xi_\phi \geq 80$ at the inflationary energy scale rather than $\xi_\phi \geq 780$ \cite{HiggsMassTopQuark}. Nonetheless, for illustrative purposes, we use $\lambda_I , g \sim 10^{-2}$ and $\xi_I \sim 10^3$ for the remainder of our analysis.

We further note that despite such large nonminimal couplings, $\xi_I \sim 10^3$, our analysis is unhindered by any potential breakdown of unitarity. The energy scale at which unitarity might be violated for Higgs inflation has occasioned a great deal of heated debate in the literature, with conflicting claims that the renormalization cut-off scale should be in the vicinity of $M_{\rm pl}$, $M_{\rm pl} / \sqrt{\xi_\phi}$, or $M_{\rm pl} / \xi_\phi$ \cite{unitarity}. Even if one adopted the most stringent of these suggested cut-off scales, $M_{\rm pl} / \xi_\phi \sim 10^{-3} \> M_{\rm pl}$, the relevant dynamics for our analysis would still occur at energy scales well below the cut-off, given the constraint $H (t_{\rm hc} ) \leq 3.7 \times 10^{-5} \> M_{\rm pl}$. (The unitarity cut-off scale in multifield models in which the nonminimal couplings $\xi_I$ are not all equal to each other has been considered in \cite{unitaritymultifield}, which likewise identify regimes of parameter space in which $\Lambda_{\rm eff}$ remains well above the energy scales and field values relevant to inflation.) Moreover, models like Higgs inflation can easily be ``unitarized" with the addition of a single heavy scalar field \cite{HiggsUnitarized}, and hence all of the following analysis could be considered the low-energy dynamics of a self-consistent effective field theory. The methods developed here may be applied to a wide class of models, including those studied in \cite{Kallosh:2013tua,Kallosh:2013daa,Kallosh:2013maa,Kallosh:2013pby}.

Finally, we note that for couplings $\lambda_I, g \sim 10^{-2}$ and $\xi_I \sim 10^3$ at high energies, the regime of weak curvature for the potential, $\kappa < 0.1$, requires that the couplings be close but not identical to each other. In particular, $\kappa \sim 0.1$ requires $g / \lambda_\phi \sim \xi_\chi / \xi_\phi \sim 1 \pm {\cal O} (10^{-5}).$ Such small differences are exactly what one would expect if the effective couplings at high energies arose from some softly broken symmetry. For example, the field $\chi$ could couple to some scalar cold dark matter (CDM) candidate (perhaps a supersymmetric partner) or to a neutrino, precisely the kinds of couplings that would be required if the primordial isocurvature perturbations were to survive to late times and get imprinted in the CMB \cite{CMBIsocurvature}. In that case, corrections to the $\beta$ functions for the renormalization-group flow of the couplings $\lambda_\chi$ and $\xi_\chi$ would appear of the form $g_{X}^2 / 16 \pi^2$ \cite{Buchbinder,SUSYsoft}, where $g_{ X}$ is the coupling of $\chi$ to the new field. For reasonable values of $g_X \sim 10^{-1} - 10^{-2}$, such additional terms could easily account for the small but non-zero differences among couplings at the inflationary energy scale. 

\subsection{Dynamics and Transfer Functions}
\label{tf}
When we vary the Einstein-frame action with respect to the fields $\phi^I$, we get the equations of motion, which may be written
\begin{equation}
\Box \phi^I + \Gamma^I_{JK} \partial_{\mu} \phi^J \partial^{\mu} \phi^K - \mathcal{G}^{IJ}V_{, K} =0,
\end{equation}
where $\Box \phi^I \equiv g^{\mu \nu} \phi^I_{; \mu ; \nu}$ and $\Gamma^I_{JK}$ is the field-space Christoffel symbol. 

We further expand each scalar field to first order in perturbations about its classical background value,
\begin{equation}
\phi^I(x^{\mu}) = \varphi^I(t) + \delta \phi^I(x^{\mu})
\end{equation}
and we consider scalar perturbations to the spacetime metric (which we assume to be a spatially flat Friedmann-Robertson-Walker metric) to first order:
\begin{equation}
\begin{split}
ds^2 &= - (1+ 2A) dt^2 + 2a(t) (\partial_i B) dx^i dt +\\  &\quad \quad a(t)^2 [(1- 2\psi) \delta_{ij} + 2 \partial_i \partial_j E] dx^i dx^j,
\end{split}
\end{equation}
where $a(t)$ is the scale factor and $A$, $B$, $\psi$ and $E$ are the scalar gauge degrees of freedom. 

Under this expansion, the full equations of motion separate into background and first-order equations. The background equations are given by 
\begin{equation}
\mathcal{D}_t \dot{\varphi}^I + 3 H \dot{\varphi}^I + \mathcal{G}^{I J} V_{, J} = 0,
\label{varphieom}
\end{equation}
where ${\cal D}_J A^I \equiv \partial_J  A^I + \Gamma^I_{\> JK} A^K$ for an arbitrary vector, $A^I$, on the field-space manifold; $\mathcal{D}_t A^I \equiv \dot{\varphi}^J \mathcal{D}_J A^I$ is a directional derivative; and  $H \equiv \dot{a} / a$ is the Hubble parameter. The $00$ and $0i$ components of the background-order Einstein equations yield:
\begin{equation}
\begin{aligned}
&H^2 = \frac{1}{3 M_{\rm pl}^2} \left[ \frac{1}{2} \mathcal{G}_{IJ} \dot{\varphi}^I \dot{\varphi}^J + V(\varphi^I) \right] \\
& \dot{H} = - \frac{1}{2M_{\rm pl}^2} \mathcal{G}_{IJ} \dot{\varphi}^I \dot{\varphi}^J.
\end{aligned}
\end{equation}
Using the covariant formalism of \cite{KMS}, we find the equations of motion for the perturbations,
\begin{equation}
\label{perteom}
\begin{split}
& \mathcal{D}_t^2 Q^I + 3 H \mathcal{D}_t Q^I + \\ &\quad  \left[\frac{k^2}{a^2} \delta^I_J + \mathcal{M}^I_{\> J} - \frac{1}{M_{\rm pl}^2 a^3} \mathcal{D}_t \left(\frac{a^3}{H} \dot{\varphi}^I \dot{\varphi}_J\right) \right] Q^J = 0,
\end{split}
\end{equation}
where $Q^I$ is the gauge-invariant Mukhanov-Sasaki variable
\begin{equation}
Q^I = \mathcal{Q}^I + \frac{\dot{\varphi}^I}{H} \psi,
\end{equation}
and $\mathcal{Q}^I$ is a covariant fluctuation vector that reduces to $\delta \phi^I$ to first order in the fluctuations. Additionally, $\mathcal{M}^I_{\> J}$ is the effective mass-squared matrix given by 
\begin{equation}
\label{mass}
\mathcal{M}^I_{\> J} \equiv \mathcal{G}^{IK} \mathcal{D}_J \mathcal{D}_K V - \mathcal{R}^I_{LMJ} \dot{\varphi}^L \dot{\varphi}^M, 
\end{equation}
where $\mathcal{R}^I_{LMJ} $ is the field-space Riemann tensor.

The degrees of freedom of the system may be decomposed into adiabatic and entropic (or isocurvature) by introducing the magnitude of the background fields' velocity vector,
\begin{equation}
\dot{\sigma} \equiv |\dot{\varphi}^I| = \sqrt{\mathcal{G}_{IJ} \dot{\varphi}^I \dot{\varphi}^J},
\end{equation}
with which we may define the unit vector
\begin{equation}
\hat{\sigma}^I \equiv \frac{\dot{\varphi}^I}{\dot{\sigma}} 
\end{equation}
which points along the fields' motion. Another important dynamical quantity is the turn-rate of the background fields, given by 
\begin{equation}
\omega^I = \mathcal{D}_t \hat{\sigma}^I ,
\end{equation}
with which we may construct another important unit vector,
\begin{equation}
\hat{s}^I \equiv \frac{\omega^I}{\omega} ,
\end{equation}
where $\omega = \abs{\omega^I}$. The vector $\hat{s}^I$ points perpendicular to the fields' motion, $\hat{s}^I \hat{\sigma}_I = 0$. The unit vectors $\hat{\sigma}^I$ and $\hat{s}^I$ effectively act like projection vectors, with which we may decompose any vector into adiabatic and entropic components. In particular, we may decompose the vector of fluctuations $Q^I$,
\begin{equation}
\begin{aligned}
& Q_{\sigma} \equiv \hat{\sigma}_I Q^I \\
& Q_s \equiv \hat{s}_I Q^I ,
\end{aligned}
\end{equation}
in terms of which Eq. (\ref{perteom}) separates into two equations of motion:
\begin{equation}
\label{adi}
\begin{split}
\ddot{Q}_{\sigma} &+ 3 H \dot{Q}_{\sigma} + \left[ \frac{k^2}{a^2} + \mathcal{M}_{\sigma \sigma} - \omega^2 - \frac{1}{M_{\text{\rm pl}}^2 a^3} \frac{d}{dt} \left(\frac{a^3 \dot{\sigma}^2}{H}\right)\right] Q_{\sigma}\\ 
& = 2\, \frac{d}{dt} \left(\omega\, Q_s\right) - 2 \left( \frac{V_{,\sigma}}{\dot{\sigma}} + \frac{\dot{H}}{H} \right) \omega\, Q_s ,
\end{split}
\end{equation}
\begin{equation}
\label{ent}
\ddot{Q}_s + 3 H \dot{Q}_s + \left[ \frac{k^2}{a^2} + \mathcal{M}_{ss} + 3 \omega^2 \right] Q_s = 4 M_{\text{\rm pl}}^2 \,\frac{\omega}{\dot{\sigma}} \frac{ k^2}{a^2} \Psi,
\end{equation}
where $\Psi$ is the gauge-invariant Bardeen potential \cite{InflationMPertsReviews},
\begin{equation}
\Psi \equiv \psi + a^2 H \left( \dot{E} - \frac{B}{a}\right),
\end{equation}
and where $ \mathcal{M}_{\sigma \sigma}$ and $ \mathcal{M}_{ss}$ are the adiabatic and entropic projections of the mass-squared matrix, $ \mathcal{M}^I_{\> J}$ from (\ref{mass}). More explicitly,
\begin{equation}
\begin{aligned}
&\mathcal{M}_{\sigma \sigma} = \hat{\sigma}_I \hat{\sigma}^J \mathcal{M}^I_{\> J} \\
& \mathcal{M}_{ss} = \hat{s}_I \hat{s}^J \mathcal{M}^I_{\> J}.
\end{aligned}\end{equation}
As Eqs. (\ref{adi}) and (\ref{ent}) make clear, the entropy perturbations will source the adiabatic perturbations but not the other way around, contingent on the turn-rate $\omega$ being nonzero. We also note that the entropy perturbations have an effective mass-squared of 
\begin{equation}
\label{eqn:mass}
\mu^2_s = \mathcal{M}_{ss} + 3 \omega^2.
\end{equation}

In the usual fashion \cite{InflationMPertsReviews}, we may construct the gauge-invariant curvature perturbation, 
\begin{equation}
\mathcal{R}_c \equiv \psi - \frac{H}{(\rho + p)} \delta q
\end{equation}
where $\rho$ and $p$ are the background-order density and pressure and $\delta q$ is the energy-density flux of the perturbed fluid. In terms of our projected perturbations, we find \cite{KMS} 
\begin{equation}
\mathcal{R}_c =  \frac{H}{\dot{\sigma}} Q_{\sigma}.
\end{equation}
Analogously, we may define a normalized entropy (or isocurvature) perturbation as \cite{InflationMPertsReviews,BassettWandsTurn,WandsBartolo,WandsReview,PTGeometric,KMS}
\begin{equation}
\mathcal{S} \equiv \frac{H}{\dot{\sigma}} Q_s.
\end{equation}

In the long-wavelength limit, the coupled perturbations obey relations of the form \cite{InflationMPertsReviews,BassettWandsTurn,WandsBartolo,WandsReview,PTGeometric,KMS}:
\begin{equation}
\begin{aligned}
&\dot{\mathcal{R}_c} \simeq \alpha H \mathcal{S}\\
&\dot{ \mathcal{S}} \simeq \beta H \mathcal{S},
\end{aligned}
\end{equation}
which allows us to write the transfer functions as
\begin{equation}
\begin{aligned}
\label{trans}
& T_{\mathcal{RS}} (t_{\rm hc}, t) = \int_{t_{\rm hc}}^t dt'\, \alpha(t')\, H(t') \,T_{\mathcal{SS}}(t_{\rm hc}, t') \\
& T_{\mathcal{SS}} (t_{\rm hc}, t) = \mathrm{exp} \left[ \int_{t_{\rm hc}}^t dt'~ \beta(t')\, H(t') \right],
\end{aligned}
\end{equation}
where $t_{\rm hc}$ is the time when a fiducial scale of interest first crosses the Hubble radius during inflation, $k_{\rm hc} = a (t_{\rm hc}) H (t_{\rm hc})$. We find \cite{KMS}
\begin{equation}
\begin{aligned}
& \alpha = \frac{2 \omega}{H}\\
&\beta = -2 \epsilon - \eta_{ss} + \eta_{\sigma \sigma} - \frac{4 \omega^2}{3 H^2},
\end{aligned}
\label{eqn:albet}
\end{equation}
where $\epsilon$, $\eta_{\sigma \sigma}$, and $\eta_{ss}$ are given by
\begin{equation}
\begin{aligned}
&\epsilon \equiv -\frac{\dot{H}}{H^2} \\
&\eta_{\sigma \sigma} \equiv \frac{M_{\text{\rm pl}}^2 \,\mathcal{M}_{\sigma \sigma}}{V} \\
&\eta_{ss} \equiv \frac{M_{\text{\rm pl}}^2 \,\mathcal{M}_{s s}}{V} .
\end{aligned}
\label{srdef}
\end{equation}
The first two quantities function like the familiar slow-roll parameters from single-field inflation: $\eta_{\sigma\sigma} = 1$ marks the end of the fields' slow-roll evolution, after which $\ddot{\sigma} \sim H \dot{\sigma}$, while $\epsilon = 1$ marks the end of inflation ($\ddot{a} = 0$ for $\epsilon = 1$). The third quantity, $\eta_{ss}$, is related to the effective mass of the isocurvature perturbations, and need not remain small during inflation.

Using the transfer functions, we may relate the power spectra at $t_{\rm hc}$ to spectra at later times. In the regime of interest, for late times and long wavelengths, we have
\begin{equation}
\begin{aligned}
&\mathcal{P}_{\mathcal{R}}(k) = \mathcal{P}_{\mathcal{R}}(k_{\rm hc}) \left[1 + T_{\mathcal{RS}}^2 (t_{\rm hc}, t)\right]\\
&\mathcal{P}_{\mathcal{S}}(k) = \mathcal{P}_{\mathcal{R}}(k_{\rm hc})\, T_{\mathcal{SS}}^2(t_{\rm hc}, t).
\end{aligned}
\end{equation}
Ultimately, we may use $T_{\mathcal{RS}}$ and $T_{\mathcal{SS}}$ to calculate the isocurvature fraction, 
\begin{equation}
\label{biso}
\beta_{\text{\rm iso}}\equiv \frac{\mathcal{P}_{\mathcal{S}}}{\mathcal{P}_{\mathcal{S}} + \mathcal{P}_{\mathcal{R}}} = \frac{T_{\mathcal{SS}}^2}{T_{\mathcal{SS}}^2 + T_{\mathcal{RS}}^2 + 1},
\end{equation}
which may be compared to recent observables reported by the \emph{Planck} collaboration. 

\begin{figure}[htb]
\includegraphics[width = 0.45\textwidth]{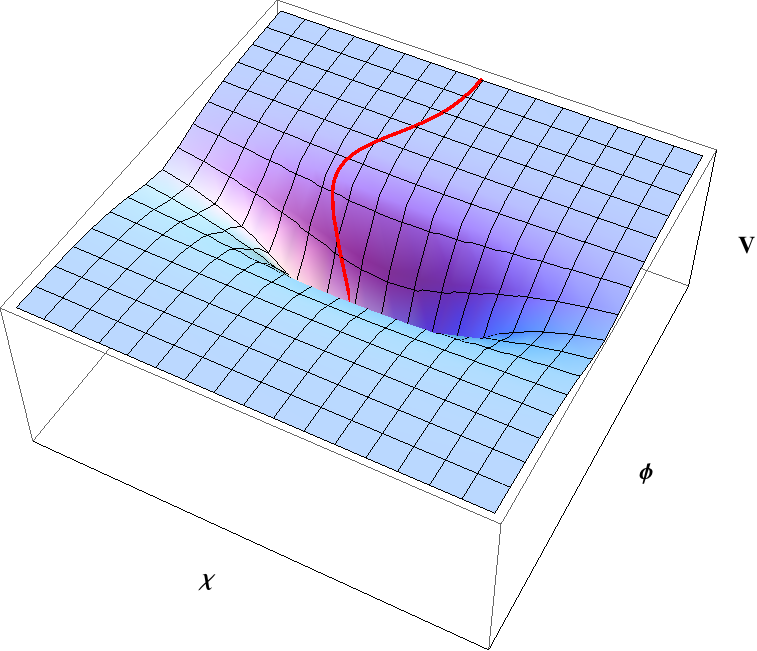}
\caption{The fields' trajectory (red) superimposed upon the effective potential in the Einstein frame, $V$, with couplings $\xi_\phi = 1000$, $\xi_\chi = 1000.015$, $\lambda_\phi = \lambda_\chi = g = 0.01$, and initial conditions $\phi_0 = 0.35$, $\chi_0 = 8.1 \times 10^{-4}$, $\dot{\phi}_0 = \dot{\chi}_0 = 0$, in units of $M_{\rm pl}$. }
\label{fig:ev}
\end{figure}

\begin{figure}
\includegraphics[width=0.45 \textwidth]{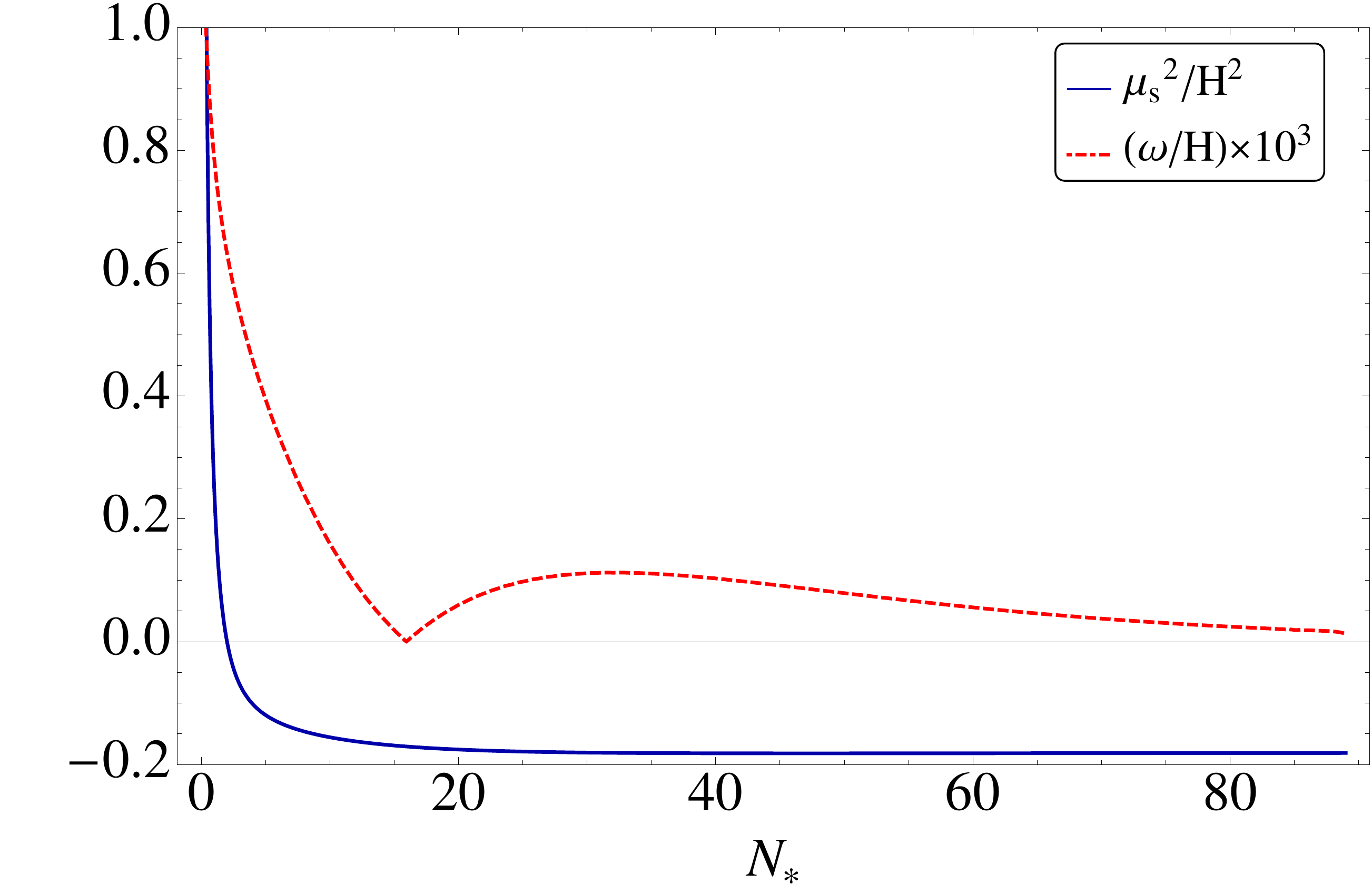}
\caption{ The mass of the isocurvature modes, $\mu_s^2 / H^2$ (blue, solid), and the turn rate, $(\omega / H) \times 10^3$ (red, dotted), versus $e$-folds from the end of inflation, $N_*$, for the trajectory shown in Fig. \ref{fig:ev}. Note that while the fields ride along the ridge, the isocurvature modes are tachyonic, $\mu_s^2 < 0$, leading to an amplification of isocurvature perturbations. The mass $\mu_s^2$ becomes large and positive once the fields roll off the ridge, suppressing further growth of isocurvature modes. }
\label{fig:musomega}
\end{figure}

An example of the fields' trajectory of interest is shown in Fig. \ref{fig:ev}. As shown in Fig. \ref{fig:musomega}, while the fields evolve near the top of the ridge, the isocurvature modes are tachyonic, $\mu_s^2 < 0$, leading to the rapid amplification of isocurvature modes. When the turn-rate is nonzero, $\omega \neq 0$, the growth of $Q_s$ can transfer power to the adiabatic perturbations, $Q_\sigma$. If $T_{\cal RS}$ grows too large from this transfer, then predictions for observable quantities such as $n_s$ can get pulled out of agreement with present observations, as shown in the intermediate region of Fig. \ref{fig:nswide} and developed in more detail in Section \ref{results}. On the other hand, growth of $Q_s$ is strongly suppressed when fields evolve in a valley, since $\mu_s^2 /H^2 \gg 1$. In order to produce an appropriate fraction of isocurvature perturbations while also keeping observables such as $n_s$ close to their measured values, one therefore needs field trajectories that stay on a ridge for a significant number of $e$-folds and have only a modest turn-rate so as not to transfer too much power to the adiabatic modes. This may be accomplished in the regime of weak curvature, $\kappa \ll 1$.

\section{Trajectories of Interest}
\label{traj}

\subsection{Geometry of the Potential}

As just noted, significant growth of isocurvature perturbations occurs when $\mu_s^2 < 0$, when the fields begin near the top of a ridge. If the fields start in a valley, or if the curvature near the top of the ridge is large enough ($\kappa \gg 1$) so that the fields rapidly fall into a valley, then the system quickly relaxes to the single-field attractor found in \cite{KS}, for which $\beta_{\rm iso} \rightarrow 0$. To understand the implications for quantities such as $\beta_{\rm iso}$, it is therefore important to understand the geomtery of the potential. This may be accomplished by working with the field-space coordinates $r$ and $\theta$, defined via
\beq
\phi = r \cos \theta~,~\chi = r \sin \theta .
\eeq
(The parameter $\theta$ was labeled $\gamma$ in \cite{GKS}.) Inflation in these models occurs for $\xi_\phi \phi^2 + \xi_\chi \chi^2 \gg M_{\rm pl}^2$ \cite{KMS}. That limit corresponds to taking $r \to \infty$, for which the potential becomes 
\beq
V_{r\to \infty} (\theta) =  {M_{\rm pl}^4 \over 4} ~   \frac{2 g \cos^2 \theta  \sin^2 \theta +  \lambda_\phi \cos^4 \theta + \lambda_\chi \sin^4 \theta }{\left(\xi_\phi  \cos^2 \theta +\xi_\chi  \sin^2 \theta \right)^2} .
\eeq
We further note that for our choice of potential in Eq. (\ref{VEpot}), $V (\phi, \chi)$ has two discrete symmetries, $\phi \to - \phi$ and $\chi \to - \chi$. This means that we may restrict our attention to only one quarter of the $\phi - \chi$ plane. We choose $\phi > 0$ and $\chi > 0$ without loss of generality. 

The extrema (ridges and valleys) are those places where $V_{, \theta} =0$, which formally has three solutions for $0<\theta<\pi/2$ and $r \rightarrow \infty$:
\beq
\theta_1 = 0 , \>\> \theta_2 = \frac{\pi}{2} , \>\> \theta_3 = {\rm cos}^{-1} \left[ \frac{ \sqrt{\Lambda_\chi}}{\sqrt{\Lambda_\phi + \Lambda_\chi } } \right] ,
\label{threethetas}
\eeq
where we have defined the convenient combinations
\beqn
\begin{split}
\Lambda_\phi  &\equiv \lambda_\phi \xi_\chi - g \xi_\phi \\
\Lambda_\chi &\equiv \lambda_\chi \xi_\phi - g \xi_\chi .
\end{split}
\label{Lambdadef}
\eeqn
In order for $\theta_3$ to be a real angle (between $0$ and $\pi/2$), the argument of the inverse cosine in Eq. (\ref{threethetas}) must be real and bounded by $0$ and $1$. If $\Lambda_\chi$ and $\Lambda_\phi$ have the same sign, both conditions are automatically satisfied. If $\Lambda_\chi$ and $\Lambda_\phi$ have different signs then the argument may be either imaginary or larger than $1$, in which case there is no real solution $\theta_3$. If both $\Lambda_\chi$ and $\Lambda_\phi$ have the same sign, the limiting cases are: for $\Lambda_\chi \gg \Lambda_\phi$, then $\theta_3 \to 0$, and for $\Lambda_\chi \ll \Lambda_\phi$ then $\theta_3 \to \pi/2$.

\begin{figure}
\includegraphics[width = 0.48\textwidth]{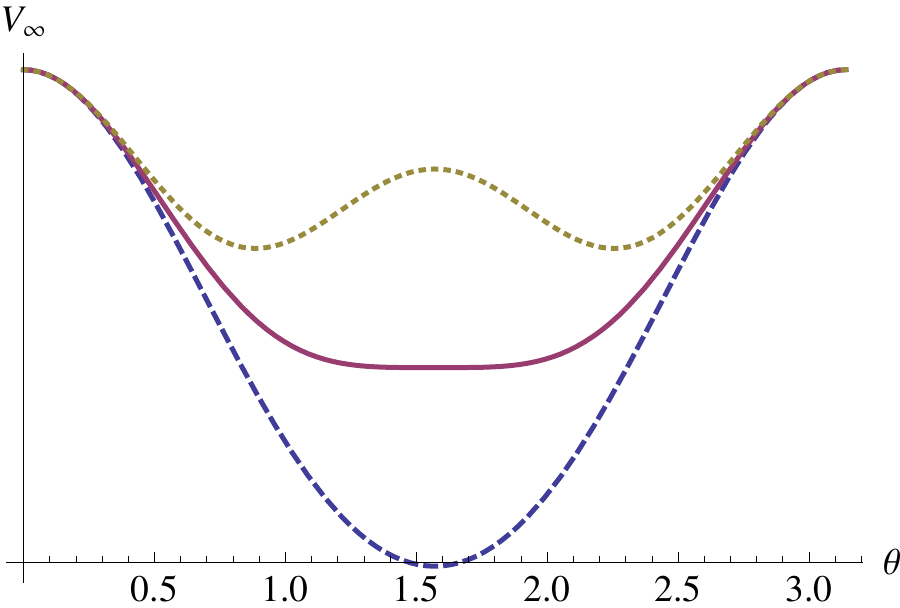}
\caption{The asymptotic value $r \rightarrow \infty$ for three potentials with $\Lambda_\chi =-0.001$ (blue dashed), $\Lambda_\chi=0$ (red solid), and $\Lambda_\chi=0.001$ (yellow dotted), as a function of the angle $\theta = \arctan(\chi / \phi)$. For all three cases, $\Lambda_\phi=0.0015$, $\xi_\phi = \xi_\chi = 1000$, and $\lambda_\phi = 0.01$.}
\label{fig:varLx}
\end{figure}

In each quarter of the $\phi - \chi$ plane, we therefore have either two or three extrema, as shown in Fig. \ref{fig:varLx}. Because of the mean-value theorem, two ridges must be separated by a valley and vice versa. If $\Lambda_\chi$ and $\Lambda_\phi$ have opposite signs, there are only two extrema, one valley and one ridge. This was the case for the parameters studied in \cite{KMS}. If $\Lambda_\phi$ and $\Lambda_\chi$ have the same sign, then there is a third extremum (either two ridges and one valley or two valleys and one ridge) within each quarter plane. In the case of two ridges, their asymptotic heights are
\beqn
\begin{split}
V_{r \rightarrow \infty} (\theta_1) &=  \frac{\lambda_\phi M_{\rm pl}^4}{4 \xi_\phi^2 } , \\
V_{r \rightarrow \infty} (\theta_2) &=  \frac{ \lambda_\chi M_{\rm pl}^4 }{4\xi_\chi^2 } ,
\end{split}
\label{Vridges}
\eeqn
and the valley lies along the direction $\theta_3$. In the limit $r \rightarrow \infty$, the curvature of the potential at each of these extrema is given by
\beqn
\begin{split}
V_{, \theta\theta} \vert_{\theta = 0} &= - \frac{ \Lambda_\phi M_{\rm pl}^4}{\xi_\phi^3} , \>\> V_{, \theta\theta} \vert_{\theta = \pi / 2} = - \frac{ \Lambda_\chi M_{\rm pl}^4}{\xi_\chi^3 } , \\
V_{, \theta\theta} \vert_{\theta = \theta_3} &=  \frac{ 2 \Lambda_\chi \Lambda_\phi ( \Lambda_\phi + \Lambda_\chi )^2 M_{\rm pl}^4 }{ ( \xi_\chi \Lambda_\phi + \xi_\chi \Lambda_\chi )^3 } .
\end{split}
\label{curvatures}
\eeqn

In this section we have ignored the curvature of the field-space manifold, since for large field values the manifold is close to flat \cite{KMS}, and hence ordinary and covariant derivatives nearly coincide. We demonstrate in Appendix B that the classification of local curvature introduced here holds generally for the dynamics relevant to inflation, even when one takes into account the nontrivial field-space manifold.

\subsection{Linearized Dynamics}

In this section we will examine trajectories for which $\omega$ is small but nonzero: small enough so that the isocurvature perturbations do no transfer all their energy away to the adiabatic modes, but large enough so that genuine multifield effects (such as $\beta_{\rm iso} \neq 0$) persist rather than relaxing to effectively single-field evolution.

We focus on situations in which inflation begins near the top of a ridge of the potential, with $\phi_0$ large and both $\chi_0$ and $\dot{\chi}_0$ small. Trajectories for which the fields remain near the top of the ridge for a substantial number of $e$-folds will produce a significant amplification of isocurvature modes, since $\mu_s^2 < 0$ near the top of the ridge and hence the isocurvature perturbations grow via tachyonic instability. From a model-building perspective it is easy to motivate such initial conditions by postulating a waterfall transition, similar to hybrid inflation scenarios \cite{hybrid}, that pins the $\chi$ field exactly on the ridge. Anything from a small tilt of the potential to quantum fluctuations would then nudge the field off-center. 

With $\chi_0$ small, sufficient inflation requires $\xi_\phi \phi_0^2 \gg M_{\rm pl}^2$, which is easily accomplished with sub-Planckian field values given $\xi_\phi \gg 1$. We set the scale for $\chi_0$ by imagining that $\chi$ begins exactly on top of the ridge. In the regime of weak curvature, $\kappa \ll 1$, quantum fluctuations will be of order
\beq
\left < \chi^2 \right > = {H^2 \over 2\pi} \Rightarrow \chi_{\rm rms} = {H\over \sqrt{2\pi}}
\eeq
where we take $\chi_{\rm rms} \equiv \sqrt{ \left < \chi^2 \right >} $ to be a classical estimator of the excursion of the field away from the ridge. The constraint from {\it Planck} that $H / M_{\rm pl} \leq 3.7 \times 10^{-5}$ during inflation then allows us to estimate $\chi_{\rm rms} \sim 10^{-5} \> M_{\rm pl}$ at the start of inflation. (A Gaussian wavepacket for $\chi$ will then spread as $\sqrt{N}$, where $N$ is the number of $e$-folds of inflation.) This sets a reasonable scale for $\chi_0$; we examine the dynamics of the system as we vary $\chi_0$ around $\chi_{\rm rms}$.

We may now expand the full background dynamics in the limit of small $\kappa, \chi$, and $\dot{\chi}$. The equation of motion for $\phi$, given by Eq. (\ref{varphieom}), does not include any terms linear in $\chi$ or $\dot{\chi}$, so the evolution of $\phi$ in this limit reduces to the single-field equation of motion, which reduces to 
\beq
\dot{\phi}_{\rm SR} \simeq - \frac{ \sqrt{ \lambda_\phi} \> M_{\rm pl}^3}{3 \sqrt{3} \xi_\phi^2 \phi }
\label{dotphiSR}
\eeq
in the slow-roll limit \cite{GKS}. To first approximation, the $\phi$ field rolls slowly along the top of the ridge. Upon using Eq. (\ref{Hamplitude}), we may integrate Eq. (\ref{dotphiSR}) to yield
\beq
\frac{\xi_\phi \phi_*^2}{ M_{\rm pl}^2} \simeq \frac{4}{3} N_* ,
\label{phiN}
\eeq
where $N_*$ is the number of $e$-folds from the end of inflation, and we have used $\phi (t_*) \gg \phi (t_{\rm end})$. The slow-roll parameters may then be evaluated to lowest order in $\chi$ and $\dot{\chi}$ and take the form \cite{KS}
 \beq
 \begin{split}
 \epsilon &\simeq {3\over 4 N_* ^2} \\
 \eta_{\sigma\sigma} &\simeq -{1\over N_*} \left ( 1- {3\over 4 N_*} \right ) .
 \end{split}
 \label{epsilonSR}
 \eeq
 
Expanding the equation of motion for the $\chi$ field and considering $\xi_\phi, \xi_\chi \gg 1$ we find the linearized equation of motion
\beq
\ddot \chi + 3H\dot \chi  -  {\Lambda_\phi M_{\rm pl}^2 \over \xi_\phi ^2} \chi \simeq 0 ,
\eeq
which has the simple solution
\beq
\chi(t) \simeq \chi_0 \exp \left [ \left ( -{3H\over 2} \pm \sqrt { {9H^2\over 2} +{ \Lambda_\phi M_{\rm pl}^2 \over  \xi_\phi^2 }} \right ) N (t) \right ] ,
\label{chiapprox}
\vspace{0.2cm}
\eeq
where we again used Eq. (\ref{Hamplitude}) for $H$, and $N (t) \equiv \int_{t_0}^t H dt'$ is the number of $e$-folds since the start of inflation. If we assume that $\Lambda_\phi M_{\rm pl}^2 / \xi_\phi^2 \ll 9 H^2 / 4$, which is equivalent to $\Lambda_\phi / \lambda_\phi \ll 3 /16$, then we may Taylor expand the square root in the exponent of $\chi(t)$. This is equivalent to dropping the $\ddot \chi$ term from the equation of motion. In this limit the solution becomes
\beq
\chi(t) \simeq \chi_0  e^{\kappa N (t)} ,
\label{chiapprox2} 
\eeq
where $\kappa$ is defined in Eq. (\ref{kappa}). Upon using the definition of $\Lambda_\phi$ in Eq. (\ref{Lambdadef}), we now recognize $\kappa = 4 \Lambda_\phi / \lambda_\phi$. Our approximation of neglecting $\ddot{\chi}$ thus corresponds to the limit $\kappa \ll 3/4$.

When applying our set of approximations to the isocurvature mass in Eq. (\ref{eqn:mass}), we find that the $\mathcal{M}_{ss}$ term dominates $\omega^2 / H^2$, and the behavior of $\mathcal{M}_{ss}$ in turn is dominated by $\mathcal{D}_J \mathcal{D}_K V$ rather than the term involving ${\cal R}^I_{JKL}$. Since we are projecting the mass-squared matrix orthogonal to the fields' motion, and since we are starting on a ridge along the $\phi$ direction, the derivative of $V$ that matters most to the dynamics of the system in this limit is $\mathcal{D}_{\chi\chi} V$ evaluated at small $\chi$. To second order in $\chi$, we find 
\begin{equation}
\begin{split}
{\cal D}_{\chi\chi} V &= - \frac{\Lambda_\phi M_{\text{\rm pl}}^4}{ \xi_\phi^3 \phi^2 } \\
&+ \frac{ M_{\text{\rm pl}}^6}{ \xi_\phi^3 (1 + 6 \xi_\phi ) \phi^4 } \left[ 2 \Lambda_\phi {(1 + 6 \xi_\phi )  \over \xi_\phi} - \lambda_\phi \varepsilon \right] \\
&+ \frac{ M_{\text{\rm pl}}^4 \, \chi^2}{ \xi_\phi^3 (1 + 6 \xi_\phi ) \phi^4 } \Big[ 3  (1 + 6 \xi_\phi ) \Lambda_\chi + \\
&\quad\quad\quad\quad\quad\quad\quad \quad + (1-\varepsilon) (1 + 6 \xi_\chi ) \Lambda_\phi  \\
& \quad \quad \quad\quad\quad \quad\quad \quad + 6 (1-\varepsilon) (1 + 6 \xi_\phi ) \Lambda_\phi - \Lambda_\phi \varepsilon \Big] ,
\end{split}
\label{eqn:deriv}
\end{equation}
where we have used $\Lambda_\phi$ and $\Lambda_\chi$ as given in Eq. (\ref{Lambdadef}) and also introduced
\beq
 \varepsilon \equiv {\xi_\phi - \xi_\chi \over \xi_\phi }= 1- {\xi_\chi \over \xi_\phi }.
 \label{varepsilon}
\eeq
These terms each illuminate an aspect of the geometry of the potential: as we found in Eq. (\ref{curvatures}), $ \Lambda_\phi$ and $ \Lambda_\chi$ are proportional to the curvature of the potential along the $\phi$ and $\chi$ axes respectively, and $\varepsilon$ is the ellipticity of the potential for large field values. Intuition coming from these geometric quantities motivates us to use them as a basis for determining the dynamics in our simulations. The approximations hold well for the first several $e$-folds of inflation, before the fields fall off the ridge of the potential.

\begin{figure}[htb]
\includegraphics[width = 0.4\textwidth]{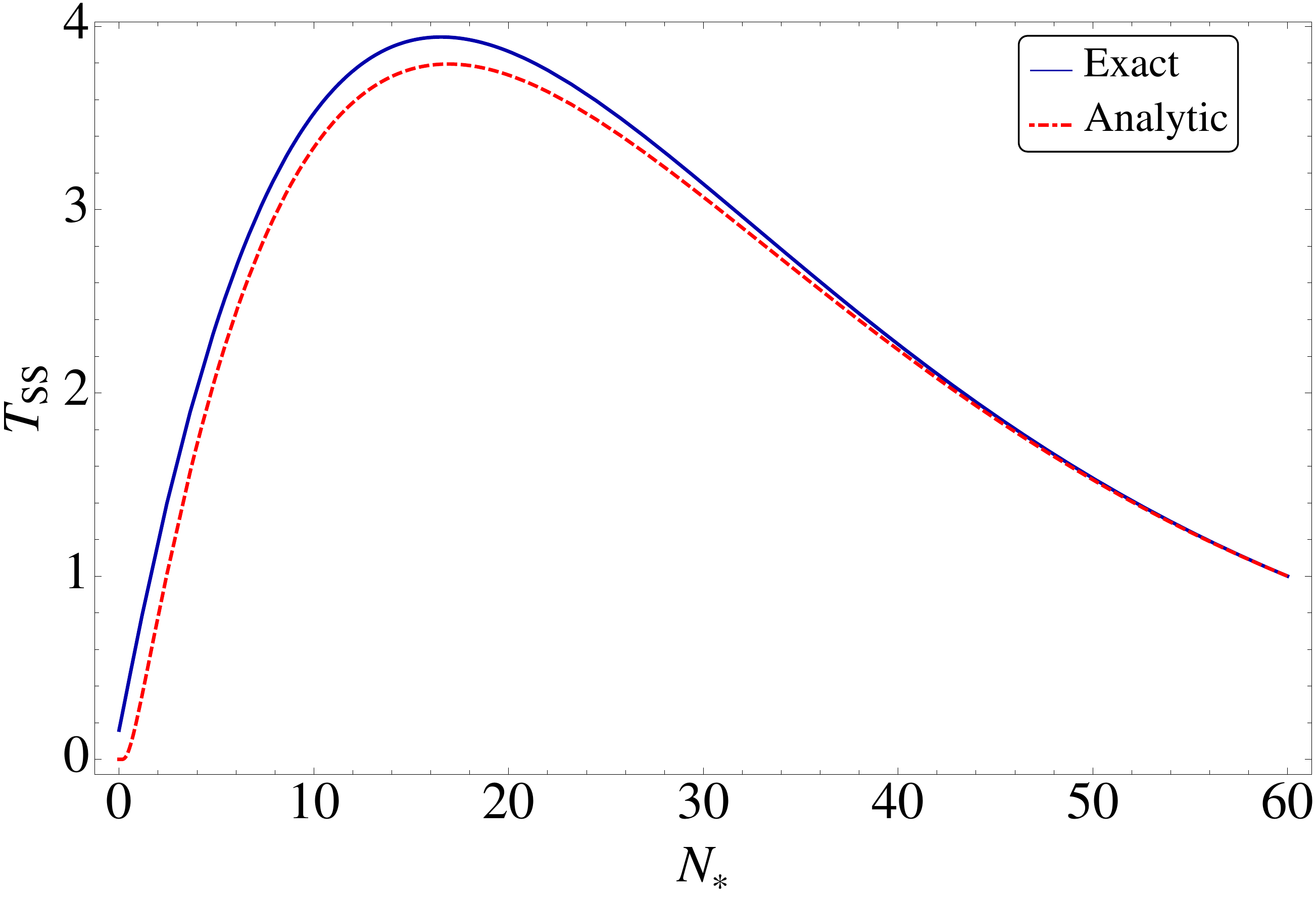}\\
\includegraphics[width = 0.4\textwidth]{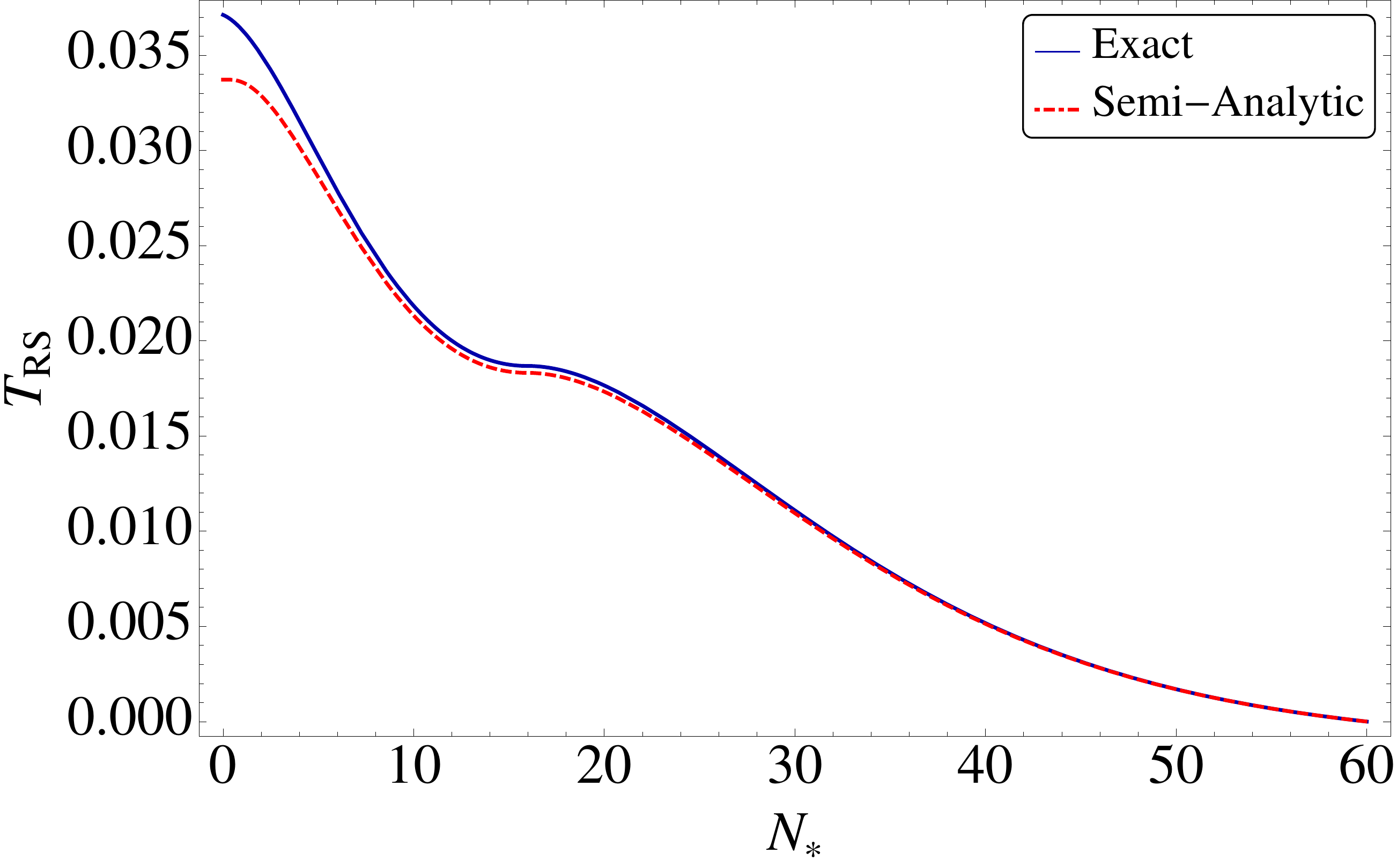} \vspace{-0.3cm}
\caption{The evolution of $T_{\mathcal{SS}}$ (top) and $T_{\mathcal{RS}}$ (bottom) using the exact and approximated expressions, for $\kappa \equiv 4 \Lambda_\phi / \lambda_\phi = 0.06$, $4 \Lambda_\chi / \lambda_\chi = - 0.06$ and $\varepsilon = -1.5 \times 10^{-5}$, with $\phi_0 = 0.35 \> M_{\rm pl}$, $\chi_0 = 8.1 \times 10^{-4} \> M_{\rm pl}$, and $\dot{\phi}_0 = \dot{\chi}_0 = 0$. We take $N_{\rm hc} = 60$ and plot $T_{\cal SS}$ and $T_{\cal RS}$ against $N_*$, the number of $e$-folds before the end of inflation. The approximation works particularly well at early times and matches the qualitative behavior of the exact numerical solution at late times.}
\label{fig:trstss}
\end{figure}

Based on our linearized approximation we may expand all kinematical quantities in power series of $\chi_0$ and $1/ N_*$. We refer to the intermediate quantities in Appendix B and report here the important quantities that characterize the generation and transfer of isocurvature perturbations. To lowest order in $\chi$ and $\dot{\chi}$, the parameter $\eta_{ss}$ defined in Eq. (\ref{srdef}) takes the form
\beq
\eta_{ss} \simeq - \kappa - \frac{3}{4 N_*} \left( \kappa + \frac{2 \varepsilon}{3} \right) + \frac{3}{8 N_*^2} \left(1 - \varepsilon \right) ,
\label{etassexpand}
\eeq
showing that to lowest order in $1/N_*$, $\eta_{ss} \sim - \kappa < 0$ and hence the isocurvature modes begin with a tachyonic mass. The quantities $\alpha$ and $\beta$ from Eq. (\ref{eqn:albet}) to first order are
\beq
\begin{split}
\alpha &\simeq \frac{ \kappa\, \chi_0\, \text{exp}\left[\kappa (N_{\rm tot} -N_*)\right]}{ \sqrt{2}\, \xi_\phi M_{\text{\rm pl}}} \sqrt{N_*} ,
\\
\beta &\simeq \kappa + \frac{1}{N_*} \left[ \frac{3 \kappa}{4} + \frac{\varepsilon}{2} - 1 \right] + \frac{1}{ N_*^2} \left[ \frac{3\varepsilon}{8} - \frac{9}{8} \right],
\end{split}
\label{alphabeta}
\eeq
where $N_{\rm tot}$ is the total number of $e$-folds of inflation. These expansions allow us to approximate the transfer function 
 $T_{\mathcal{SS}}$ of Eq. (\ref{trans}), 
\beq
\begin{split}
T_{\cal SS} &\simeq \left( \frac{ N_*}{N_{\rm hc} } \right)^{1 - {3 \kappa \over 4 } - { \varepsilon \over  2} }  \\
&\quad\quad \times \exp \left[ \kappa \left( N_{\rm hc} - N_* \right) - \frac{3}{8} \left( 3 - \varepsilon \right) \left( \frac{1}{ N_*} - \frac{1}{ N_{\rm hc} } \right) \right] ,
\end{split}
\label{TRSTSSapprox}
\eeq
where $N_{\rm hc}$ is the number of $e$-folds before the end of inflation at which Hubble crossing occurs for the fiducial scale of interest. We may then use a semi-analytic form for $T_{\cal RS}$ by putting Eq. (\ref{TRSTSSapprox}) into Eq. (\ref{trans}). This approximation is depicted in Fig. \ref{fig:trstss}. 

Our analytic approximation for $T_{\cal SS}$ vanishes identically in the limit $N_* \rightarrow 0$ (at the end of inflation), though it gives an excellent indication of the general shape of $T_{\cal SS}$ for the duration of inflation. We further note that $T_{\cal SS}$ is independent of $\chi_0$ to lowest order, while $T_{\cal RS} \propto \alpha \propto \kappa \chi_0$ and hence remains small in the limit we are considering. Thus for small $\kappa$, we expect $\beta_{\rm iso}$ to be fairly insensitive to changes in $\chi_0$.

\section{Results}
\label{results}

We want to examine how the isocurvature fraction $\beta_\text{\rm iso}$ varies as we change the shape of the potential. We are particularly interested in the dependence of $\beta_{\rm iso}$ on $\kappa$, since the leading-order contribution to the isocurvature fraction from the shape of the potential is proportional to $\kappa$. Guided by our approximations, we simulated trajectories across 1400 potentials and we show the results in Figures \ref{fig:chi} - \ref{fig:e}. The simulations were done using zero initial velocities for $\phi$ and $\chi$, and were performed using both Matlab and Mathematica, as a consistency check. We compare analytical approximations in certain regimes with our numerical findings.

As expected, we find that there is an interesting competition between the degree to which the isocurvature mass is tachyonic and the propensity of the fields to fall off the ridge. More explicitly, for small $\kappa$ we expect the fields to stay on the ridge for most of inflation with a small turn rate that transfers little power to the adiabatic modes. Therefore, in the small-$\kappa$ limit, $T_\mathcal{RS}$ remains small while $T_\mathcal{SS}$ (and hence $\beta_\text{\rm iso}$) increases exponentially with increasing $\kappa$. Indeed, all the numerical simulations show that $\beta_\text{\rm iso}$ vs. $\kappa$ increases linearly on a semilog scale for small $\kappa$. However, in the small-$\kappa$ limit, the tachyonic isocurvature mass is also small, so $\beta_\text{\rm iso}$ remains fairly small in that regime. Meanwhile, for large $\kappa$ we expect the fields to have a larger tachyonic mass while near the top of the ridge, but to roll off the ridge (and transfer significant power to the adiabatic modes) earlier in the evolution of the system. There should be an intermediate regime of $\kappa$ in which the isocurvature mass is fairly large (and tachyonic) and the fields do not fall off the ridge too early. Indeed, a ubiquitous feature of our numerical simulations is that $\beta_\text{\rm iso}$ is always maximized around $\kappa \lesssim  0.1$, regardless of the other parameters of the potential.

\subsection{Local curvature of the potential}

\begin{figure}[htb]
\includegraphics[width = 0.48\textwidth]{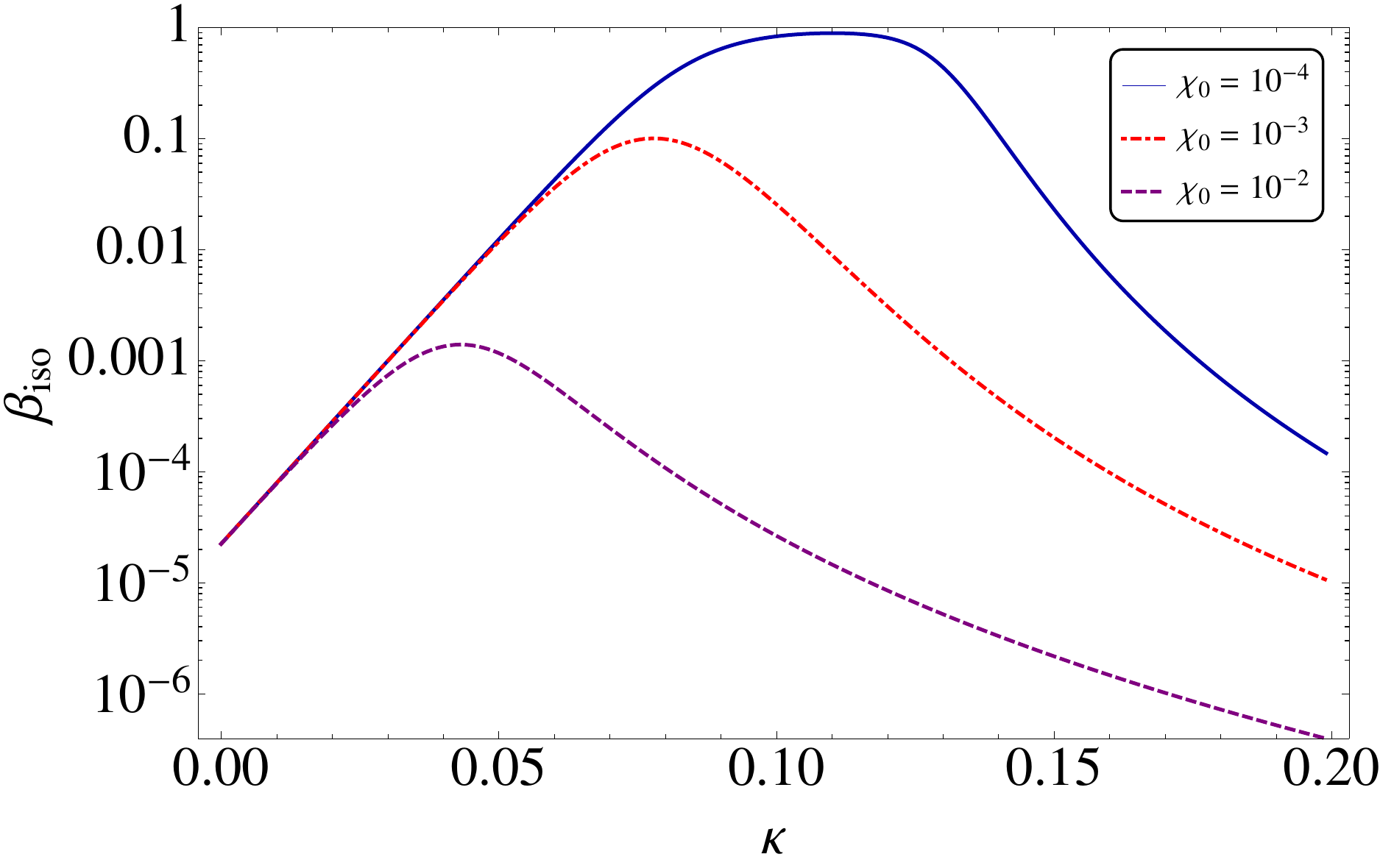}\vspace{-0.4cm}
\caption{The isocurvature fraction for different values of $\chi_0$ (in units of $M_{\rm pl}$) as a function of the curvature of the ridge, $\kappa$. All of the trajectories began at $\phi_0 = 0.3 \> M_{\rm pl}$, which yields $N_{\rm tot} = 65.7$. For these potentials, $\xi_\phi = 1000$, $\lambda_\phi = 0.01$, $\varepsilon = 0$, and $\Lambda_\chi = 0$. The trajectories that begin closest to the top of the ridge have the largest values of $\beta_\text{\rm iso}$, with some regions of parameter space nearly saturating $\beta_\text{\rm iso}$ = 1.}
\label{fig:chi}
\end{figure}

In Fig. \ref{fig:chi}, we examine the variation of $\beta_{\rm iso}$ as we change $\chi_0$ and $\kappa$.  
As expected, $\beta_\text{\rm iso}$ has no dependence on $\chi_0$ for small $\kappa$. Increasing $\kappa$ breaks the $\chi_0$ degeneracy:  the closer the fields start to the top of the ridge, the more time the fields remain near the top before rolling off the ridge and transferring power to the adiabatic modes. Just as expected, for the smallest value of $\chi_0$, we see the largest isocurvature fraction. Even for relatively large $\chi_0$, there is still a nontrivial contribution of isocurvature modes to the perturbation spectrum. Therefore, our model generically yields a large isocurvature fraction with little fine-tuning of the initial field values in the regime $\kappa \ll 1$. 

We may calculate $\beta_{\rm iso}$ for the limiting case of zero curvature, $\kappa \to 0$, the vicinity in which the curves in Fig. \ref{fig:chi}  become degenerate. Taking the limit $\kappa \to 0$ means essentially reverting to a Higgs-like case, a fully $SO(2)$ symmetric potential with no turning of the trajectory in field space \cite{GKS}. As expected, our approximate expression in Eq. (\ref{TRSTSSapprox}) for $T_{\cal RS} \to 0$ in the limit $\kappa \to 0$, and hence we need only consider $T_{\cal SS}$.

As noted above, our approximate expression for $T_{\cal SS}$ in Eq. (\ref{TRSTSSapprox}) vanishes in the limit $N_* \to 0$. Eq. (\ref{TRSTSSapprox}) was derived for the regime in which our approximate expressions for the slow-roll parameters $\epsilon$ and $\eta_{\sigma\sigma}$ in Eq. (\ref{epsilonSR}) are reasonably accurate. Clearly the expressions in Eq. (\ref{epsilonSR}) will cease to be accurate near the end of inflation.  Indeed, taking the expressions at face value, we would expect slow roll to end ($\vert \eta_{\sigma\sigma} \vert = 1$) at $N_* = 1/2$, and inflation to end ($\epsilon = 1$) at $N_* = 2 / \sqrt{3}$, rather than at $N_* = 0$. Thus we might expect Eq. (\ref{epsilonSR}) to be reliable until around $N_* \simeq 1$, which matches the behavior we found in a previous numerical study \cite{KS}. Hence we will evaluate our analytic approximation for $T_{\cal SS}$ in Eq. (\ref{TRSTSSapprox}) between $N_{\rm hc} = 60$ and $N_* \simeq 1$, rather than all the way to $N_* \to 0$. In the limit $\kappa \to 0$ and $\varepsilon \to 0$ and using $N_* = 1$, Eq. (\ref{TRSTSSapprox}) yields
\beq
T_{\cal SS} \simeq \frac{1}{N_{\rm hc} } \exp \left[ - 9/8 \right],
\label{TSSapprox}
\eeq
upon taking $N_{\rm hc} \gg N_*$. For $N_{\rm hc} = 60$, we therefore find $T_{\cal SS} \simeq 5.4 \times 10^{-3}$, and hence $\beta_{\rm iso} \simeq 2.9 \times 10^{-5}$. This value may be compared with the exact numerical value, $\beta_{\rm iso} = 2.3 \times 10^{-5}$. Despite the severity of our approximations, our analytic expression provides an excellent guide to the behavior of the system in the limit of small $\kappa$.

\begin{figure}
\includegraphics[width = 0.45\textwidth]{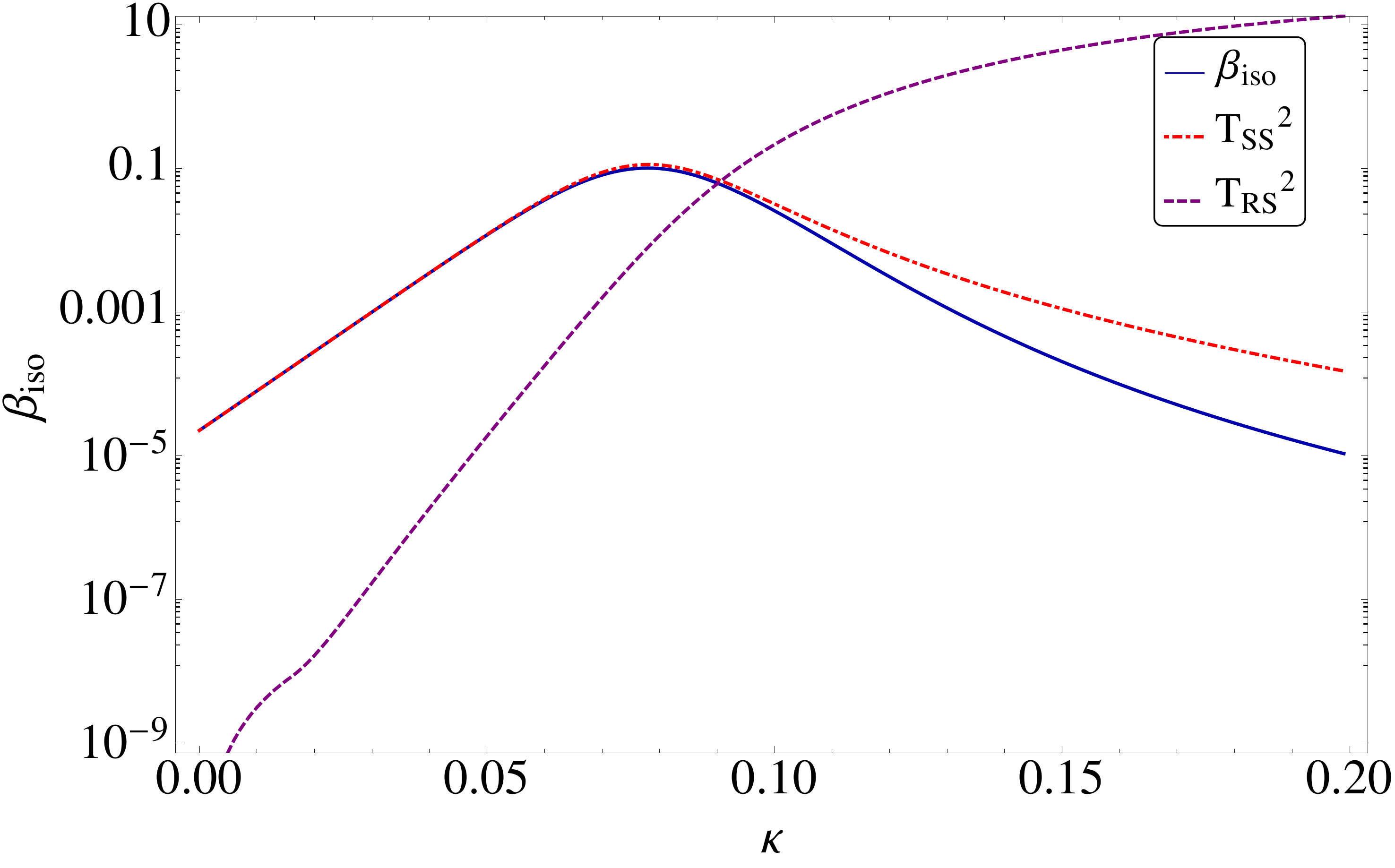}
\caption{ Contributions of $T_{\cal RS}$ and $T_{\cal SS}$ to $\beta_{\rm iso}$. The parameters used are $\phi_0 = 0.3 \> M_{\rm pl}$, $\chi_0=10^{-3} \> M_{\rm pl}$, $\xi_\phi = 10^3$, $\lambda_\phi = 0.01$, $\varepsilon = 0$ and $\Lambda_\chi = 0$. For small $\kappa$,  $\beta_{\rm iso}$ is dominated by $T_{\cal SS}$; for larger $\kappa$, $T_{\cal RS}$ becomes more important and ultimately reduces $\beta_{\rm iso}$.}
\label{fig:trstssbiso}
\end{figure}

As we increase $\kappa$, the fields roll off the ridge correspondingly earlier in their evolution. The nonzero turn-rate causes a significant transfer of power from the isocurvature modes to the adiabatic modes. As $T_{\cal RS}$ grows larger, it lowers the overall value of $\beta_{\rm iso}$. See Fig. \ref{fig:trstssbiso}.

\subsection{Global structure of the potential}

\begin{figure}
\includegraphics[width = 0.48\textwidth]{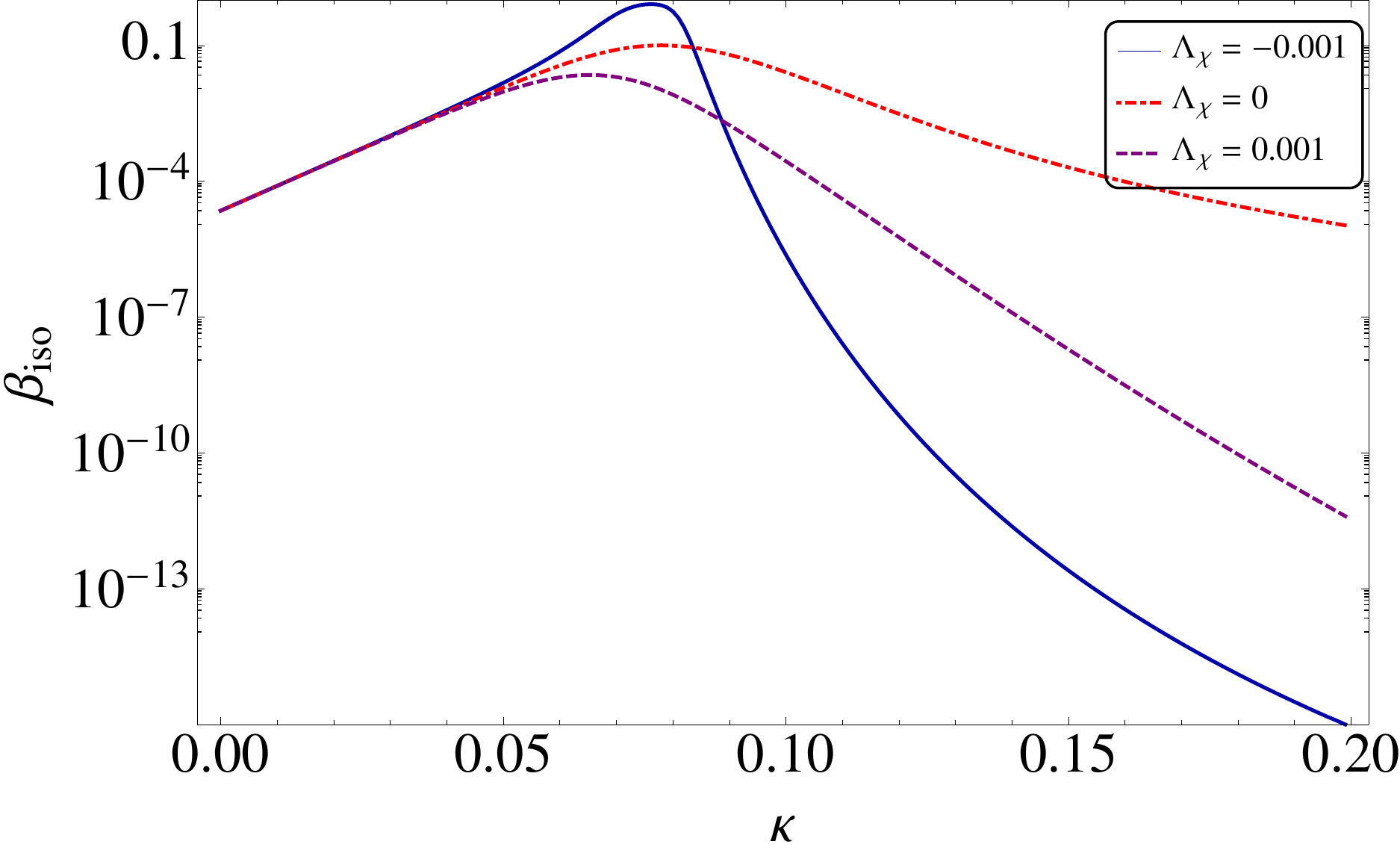}
\caption{The isocurvature fraction for different values of $\Lambda_\chi$ as a function of the curvature of the ridge, $\kappa$. All of the trajectories began at $\phi_0 = 0.3 \> M_{\rm pl}$ and $\chi_0 = 10^{-4} \> M_{\rm pl}$, yielding $N_{\rm tot}$ = 65.7. For these potentials, $\xi_\phi = 1000$, $\lambda_\phi = 0.01$, and $\varepsilon = 0$. Potentials with $\Lambda_\chi < 0$ yield the largest $\beta_\text{\rm iso}$ peaks, though in those cases $\beta_{\rm iso}$ falls fastest in the large-$\kappa$ limit due to sensitive changes in curvature along the trajectory. Meanwhile, potentials with positive $\Lambda_\chi$ suppress the maximum value of $\beta_\text{\rm iso}$ once $\kappa \gtrsim 0.1$ and local curvature becomes important. }
\label{fig:lambda}
\end{figure}

The previous discussion considered the behavior for $\Lambda_\chi = 0$. As shown in Fig. \ref{fig:varLx}, the global structure of the potential will change if $\Lambda_\chi \neq 0$. In the limit $\kappa \ll 1$, the fields never roll far from the top of the ridge along the $\chi = 0$ direction, and therefore the shape of the potential along the $\chi$ direction has no bearing on $\beta_{\rm iso}$. However, large $\kappa$ breaks the degeneracy in $\Lambda_\chi$ because the fields will roll off the original ridge and probe features of the potential along the $\chi$ direction. See Fig. \ref{fig:lambda}.

In the case $\Lambda_\chi = 0$, the fields roll off the ridge and eventually land on a plain, where the isocurvature perturbations are minimally suppressed, since $\mu_s^2 \sim 0$. For $\Lambda_\chi > 0$, there is a ridge along the $\chi$ direction as well as along $\chi = 0$, which means that there must be a valley at some intermediate angle in field space. When the fields roll off the original ridge, they reach the valley in which $\mu_s^2 > 0$, and hence the isocurvature modes are more strongly suppressed than in the $\Lambda_\chi = 0$ case. 

Interesting behavior may occur for the case $\Lambda_\chi < 0$. There exists a range of $\kappa$ for which the isocurvature perturbations are more strongly amplified than a naive estimate would suggest, thanks to the late-time behavior of $\eta_{ss} \sim ( {\cal D}_{\chi \chi} V ) / V$. If the second derivative decreases more slowly than the potential itself, then the isocurvature modes may be amplified for a short time as the fields roll down the ridge. This added contribution is sufficient to increase $\beta_{\rm iso}$ compared to the cases in which $\Lambda_\chi \geq 0$. However, the effect becomes subdominant as the curvature of the original ridge, $\kappa$, is increased. For larger $\kappa$, the fields spend more time in the valley, in which the isocurvature modes are strongly suppressed.

\begin{figure}[htb]
\includegraphics[width = 0.48\textwidth]{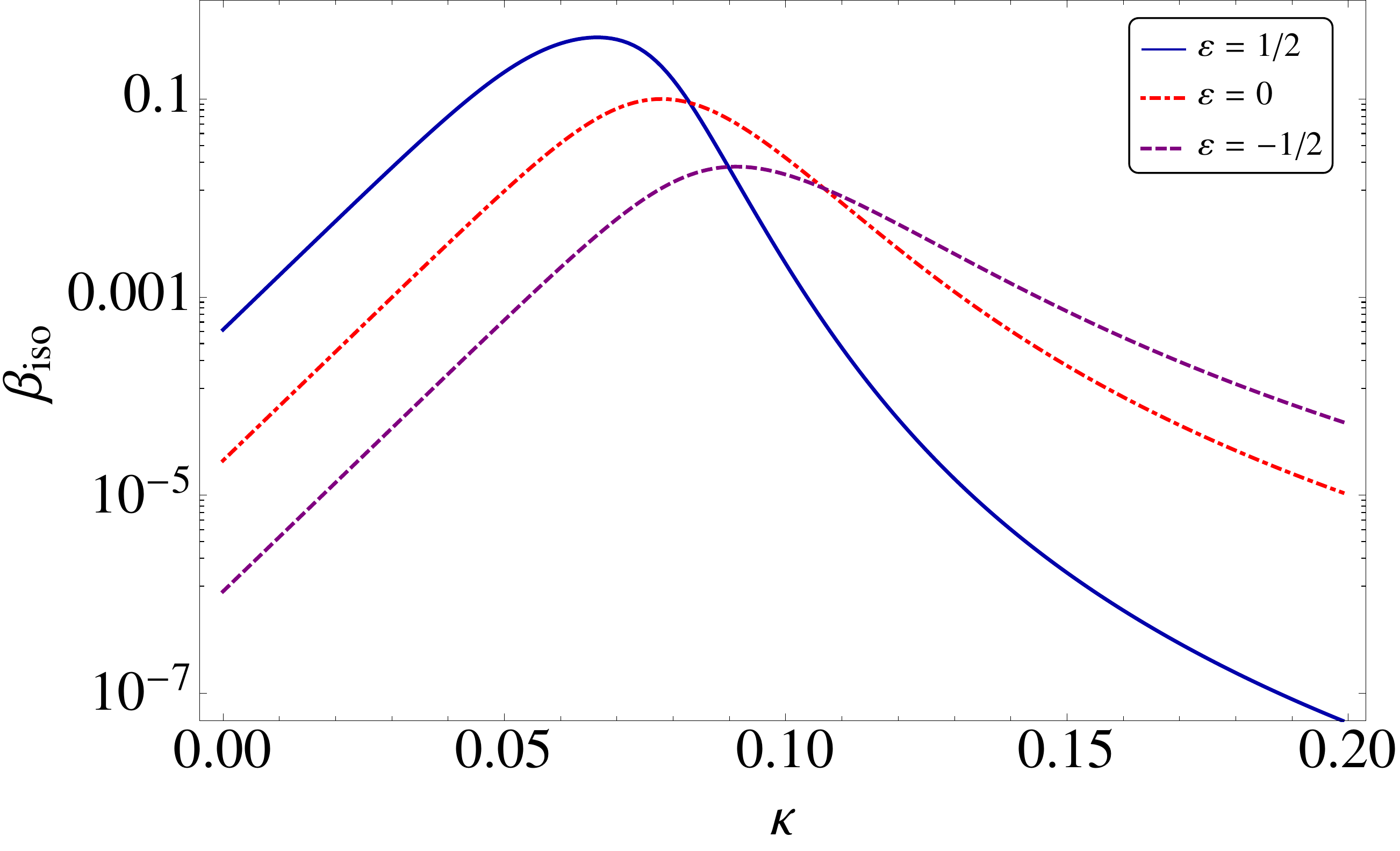}\vspace{-0.2cm}
\caption{The isocurvature fraction for different values of $\varepsilon$ as a function of the curvature of the ridge, $\kappa$. All of the trajectories began at $\chi_0 = 10^{-3} \> M_{\rm pl}$ and $\phi_0 = 0.3 \> M_{\rm pl}$, with $N_{\rm tot} = 65.7$. For these potentials, $\xi_\phi = 1000$, $\lambda_\phi = 0.01$, and $\Lambda_\chi = 0$. Here we see the competition between $\varepsilon$ setting the scale of the isocurvature mass and affecting the amount of turning in field-space.} \vspace{-0.1cm}
\label{fig:e}
\end{figure}

In Figure \ref{fig:e}, we isolate effects of $\varepsilon$ and $\kappa$ on $\beta_{\rm iso}$. From Eq. (\ref{eqn:deriv}), when $\Lambda_\phi$ is small (which implies that $\kappa$ is small), $\varepsilon$ sets the scale of the isocurvature mass. Positive $\varepsilon$ makes the isocurvature mass-squared more negative near $\kappa$ = 0, which increases the power in isocurvature modes. Conversely, negative $\varepsilon$ makes the isocurvature mass-squared less negative near $\kappa = 0$, which decreases the power in isocurvature modes. In geometrical terms, in the limit $\Lambda_\phi = \Lambda_\chi = 0$, equipotential surfaces are ellipses with eccenticity $\sqrt{\varepsilon} $ for $\varepsilon > 0$ and $ \sqrt{ \varepsilon /(\varepsilon-1)}$ for $\varepsilon < 0$. In this limit we may calculate $\beta_{\rm iso}$ exactly as we did for the case of $\varepsilon =0$.

The other effect of changing $\varepsilon$ is that it elongates the potential in either the $\phi$ or $\chi$ direction. This deformation of the potential either enhances or decreases the degree to which the fields can turn, which in turn will affect the large-$\kappa$ behavior. In particular, for $\varepsilon > 0$ the potential is elongated along the $\phi$ direction, which means that when the fields roll off the ridge, they immediately start turning and transferring power to the adiabatic modes. Conversely, for $\varepsilon < 0$ the potential is elongated along the $\chi$ direction, so once the fields fall off the ridge, they travel farther before they start turning. Therefore, in the large-$\kappa$ limit, $\beta_\text{\rm iso}$ falls off more quickly for $\varepsilon > 0$ than for $\varepsilon < 0$. 
\label{sweep}

We may use our analytic expression for $T_{\cal SS}$ in Eq. (\ref{TRSTSSapprox}) for the case in which $\kappa \to 0$ with $\varepsilon \neq 0$. We find the value of $\beta_{\rm iso} \simeq T_{\cal SS}^2$ changes by a factor of 11 when we vary $\varepsilon \pm 1/2$, while our numerical solutions in Fig. \ref{fig:e} vary by a factor of 21. Given the severity of some of our analytic approximations, this close match again seems reassuring.

\subsection{Initial Conditions}

The quantity $\beta_{\rm iso}$ varies with the fields' initial conditions as well as with the parameters of the potential. Given the form of $T_{\cal RS}$ and $T_{\cal SS}$ in Eq. (\ref{trans}), we see that the value of $\beta_{\rm iso}$ depends only on the behavior of the fields between $N_{\rm hc}$ and the end of inflation. This means that if we were to change $\phi_0$ and $\chi_0$ in such a way that the fields followed the same trajectory following $N_{\rm hc}$, the resulting values for $\beta_{\rm iso}$ would be identical. 

We have seen in Eq. (\ref{phiN}) that we may use $\phi$ as our inflationary clock, $\xi_\phi \phi_*^2 / M_{\rm pl}^2 \simeq 4 N_* / 3$, where $N_* = N_{\rm tot} - N (t)$ is the number of $e$-folds before the end of inflation. We have also seen, in Eq. (\ref{chiapprox2}), that for small $\kappa$ we may approximate $\chi (t) \simeq \chi_0 \exp [\kappa N (t) ]$. If we impose that two such trajectories cross $N_{\rm hc}$ with the same value of $\chi$, then we find
\beq
\label{eqn:contours}
\Delta(\log \chi_0) = \kappa \Delta N = - {3\over 4 } \xi_\phi \kappa~ \Delta\left ( {\phi_0 ^2 \over M_{\rm pl}^2}\right ) .
\eeq

\begin{figure}[htb]
\includegraphics[width = 0.45\textwidth]{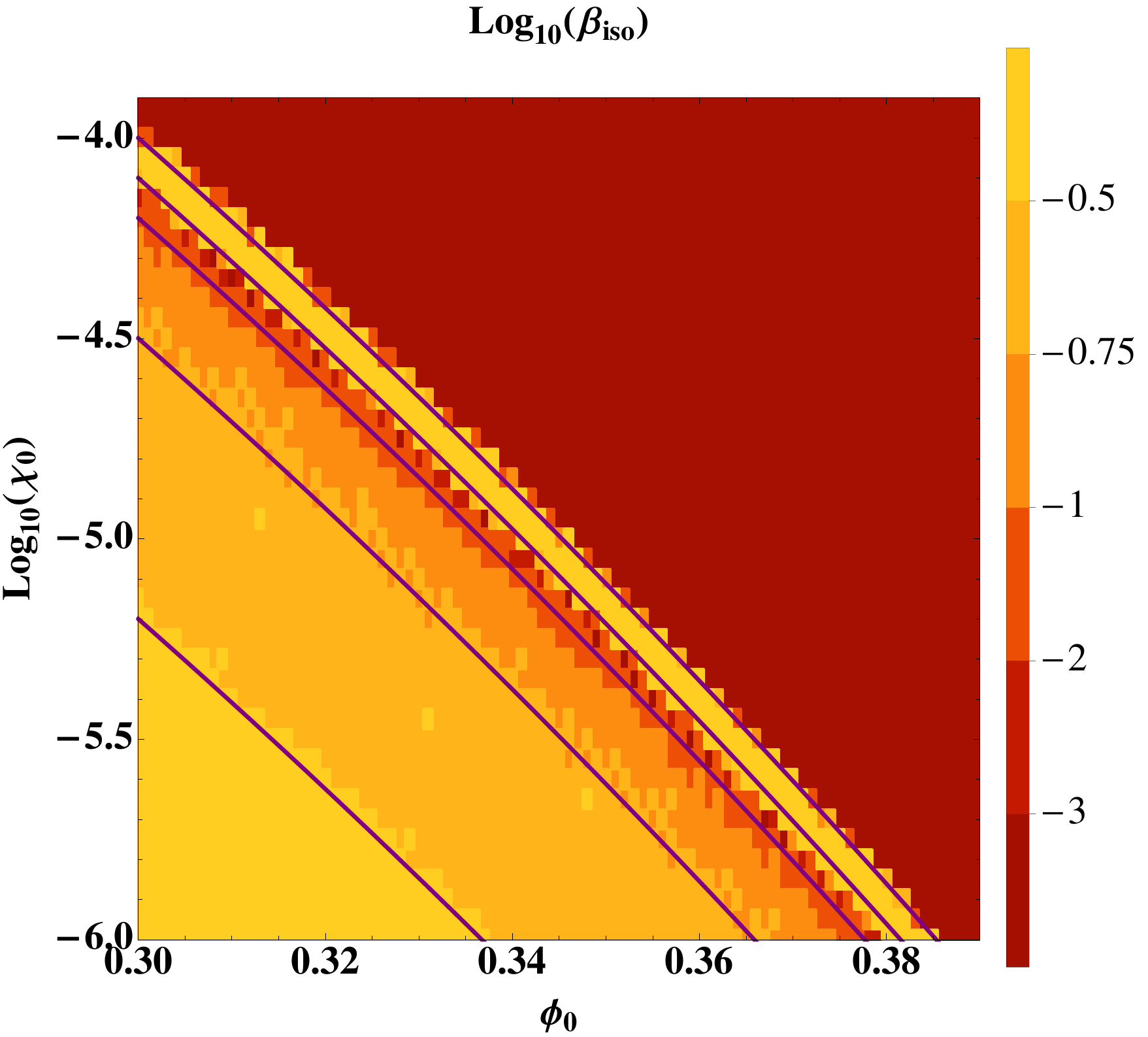}\vspace{-0.4cm}
\caption{Numerical simulations of $\beta_\text{\rm iso}$ for various initial conditions (in units of $M_{\rm pl}$). All trajectories shown here were for a potential with $\kappa = 4 \Lambda_\phi / \lambda_\phi = 0.116$, $4 \Lambda_\chi / \lambda_\chi = -160.12$, and $\varepsilon = -2.9 \times 10^{-5}$. Also shown are our analytic predictions for contours of constant $\beta_{\rm iso}$, derived from Eq. (\ref{eqn:contours}) and represented by dark, solid lines. From top right to bottom left, the contours have $\beta_\text{\rm iso} = $ 0.071, 0.307, 0.054, 0.183, and 0.355.}
\label{fig:init}
\end{figure}

We tested the approximation in Eq. (\ref{eqn:contours}) by numerically simulating over 15,000 trajectories in the same potential with different initial conditions. The numerical results are shown in Fig. \ref{fig:init}, along with our analytic predictions, from Eq. (\ref{eqn:contours}), that contours of constant $\beta_{\rm iso}$ should appear parabolic in the semilog graph. As shown in Fig. \ref{fig:init}, our analytic approximation matches the full numerical results remarkably well. We also note from Fig. \ref{fig:init} that for a given value of $\chi_0$, if we increase $\phi_0$ (thereby increasing the total duration of inflation, $N_{\rm tot}$), we will decrease $\beta_{\rm iso}$, behavior that is consistent with our approximate expressions for $T_{\cal RS}$ and $T_{\cal SS}$ in Eq. (\ref{TRSTSSapprox}).

\begin{figure}[htb]
\includegraphics[width = 0.45\textwidth]{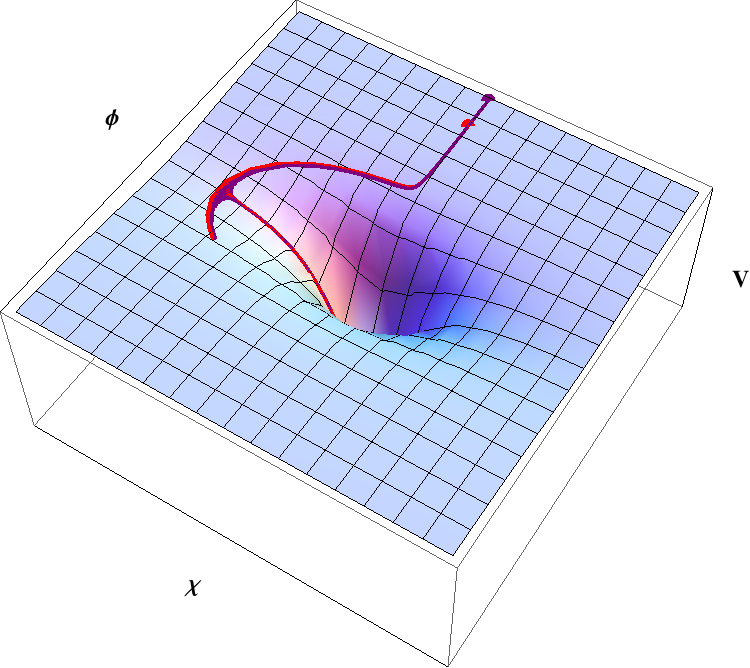}\vspace{-0.4cm}
\caption{Two trajectories from Fig. \ref{fig:init} that lie along the $\beta_\text{\rm iso}  = 0.183$ line, for $\phi_0=0.3 \> M_{\rm pl}$ and $\phi_0=  0.365 \> M_{\rm pl}$. The dots mark the fields' initial values. The two trajectories eventually become indistinguishable, and hence produce identical values of $\beta_{\rm iso}$.}
\label{fig:trajinit}
\end{figure}

\subsection{CMB observables }

Recent analyses of the {\it Planck} data for low multipoles suggests an improvement of fit between data and underlying model if one includes a substantial fraction of primordial isocurvature modes, $\beta_{\rm iso} \sim {\cal O} (0.1)$. The best fits are obtained for isocurvature perturbations with a slightly blue spectral tilt, $n_I \equiv 1 + d \ln {\cal P}_{\cal S} / d \ln k \geq 1.0$ \cite{PlanckInflation}. In the previous sections we have demonstrated that our general class of models readily produces $\beta_{\rm iso} \sim {\cal O} (0.1)$ in the regime $\kappa \lesssim 0.1$. The spectral tilt, $n_I$, for these perturbations goes as \cite{WandsBartolo,PTGeometric}
\beq
n_I = 1 - 2 \epsilon +  2 \eta_{ss} ,
\label{nI}
\eeq
where $\epsilon$ and $\eta_{ss}$ are evaluated at Hubble-crossing, $N_{\rm hc}$.  Given our expressions in Eqs. (\ref{epsilonSR}) and (\ref{etassexpand}), we then find
\beq
n_I \simeq 1 - 2 \kappa - \frac{3}{2 N_*} \left( \kappa + \frac{2 \varepsilon}{3} \right) - \frac{3}{4 N_*^2} \left( 1 + \varepsilon \right) .
\label{nI2}
\eeq
For trajectories that produce a nonzero fraction of isocurvature modes, the isocurvature perturbations are tachyonic at the time of Hubble-crossing, with $\eta_{ss} \propto {\cal M}_{ss} \sim \mu_s^2 < 0$. Hence in general we find $n_I$ will be slightly red-tilted, $n_I \leq 1$. However, in the regime of weak curvature, $\kappa \ll 1$, we may find $n_I \sim 1$. In particular, in the limit $\kappa \to 0$ and $\varepsilon \to 0$, then $n_I \to 1 - 3 / (4 N_*^2) \sim 1 - {\cal O} (10^{-4})$, effectively indistinguishable from a flat, scale-invariant spectrum. In general for $\kappa < 0.02$, we therefore expect $n_I > n_s$, where $n_s \sim 0.96$ is the spectral index for adiabatic perturbations. In that regime, the isocurvature perturbations would have a bluer spectrum than the adiabatic modes, albeit not a genuinely blue spectrum. An important test of our models will therefore be if future observations and analysis require $n_I > 1$ in order to address the present low-$\ell$ anomaly in the {\it Planck} measurements of the CMB temperature anisotropies.

Beyond $\beta_{\rm iso}$ and $n_I$, there are other important quantities that we need to address, and that can be used to distinguish between similar models: the spectral index for the adiabatic modes, $n_s$, and its running, $\alpha \equiv d n_s / d \ln k$; the tensor-to-scalar ratio, $r$; and the amplitude of primordial non-Gaussianity, $f_{\rm NL}$. As shown in \cite{KS}, in the limit of large curvature, $\kappa \gg 1$, the system quickly relaxes to the single-field attractor for which $0.960 \leq n_s \leq 0.967$, $\alpha \sim {\cal O} (10^{-4} )$, $0.0033 \leq r \leq 0.0048$, and $\vert f_{\rm NL} \vert \ll 1$. (The ranges for $n_s$ and $r$ come from considering $N_{\rm hc} = 50 - 60$.) Because the single-field attractor evolution occurs when the fields rapidly roll off a ridge and remain in a valley, in which $\mu_s^2 > 0$, the models generically predict $\beta_{\rm iso} \ll 1$ in the limit $\kappa \gg 1$ as well. Here we examine how these observables evolve in the limit of weak curvature, $\kappa \ll 1$, for which, as we have seen, the models may produce substantial $\beta_{\rm iso} \sim {\cal O} (0.1)$.

Let us start with the spectral index, $n_s$. If isocurvature modes grow and transfer substantial power to the adiabatic modes before the end of inflation, then they may affect the value of $n_s$. In particular, we have \cite{WandsBartolo,PTGeometric,KMS}
\beq
n_s = n_s(t_{\rm hc}) + {1\over H} \left[ -\alpha(t_{\rm hc} ) -\beta(t_{\rm hc} )  T_{\cal RS}  \right] \sin(2\Delta) ,
\label{nsend}
\eeq 
where
\beq
n_s (t_{\rm hc}) = 1 - 6 \epsilon + 2 \eta_{\sigma\sigma}
\label{nshc}
\eeq
and $\alpha$ and $\beta$ are given in Eq. (\ref{eqn:albet}). The angle $\Delta$ is defined via
\beq
\cos \Delta \equiv {T_{\cal RS} \over \sqrt{1+ T_{\cal RS}^2}} .
\eeq 
The turn rate $\alpha = 2 \omega / H$ is small at the moment when perturbations exit the Hubble radius, and the trigonometric factor obeys $-1\leq  \sin(2\Delta) \leq 1$. We also have $\beta \simeq \kappa + {\cal O} (N_*^{-1})$ at early times, from Eq. (\ref{alphabeta}). Hence we see that $T_{\cal RS}$ must be significant in order to cause a substantial change in $n_s$ compared to the value at Hubble crossing, $n_s (t_{\rm hc})$. Yet we found in Fig. \ref{fig:trstssbiso} that $T_{\cal RS}$ grows large after $\beta_{\rm iso}$ has reached its maximum value. We therefore expect $n_s$ to be equal to its value in the single-field attractor for $\kappa \lesssim 0.1$. 

\begin{figure}
\includegraphics[width = 0.45\textwidth]{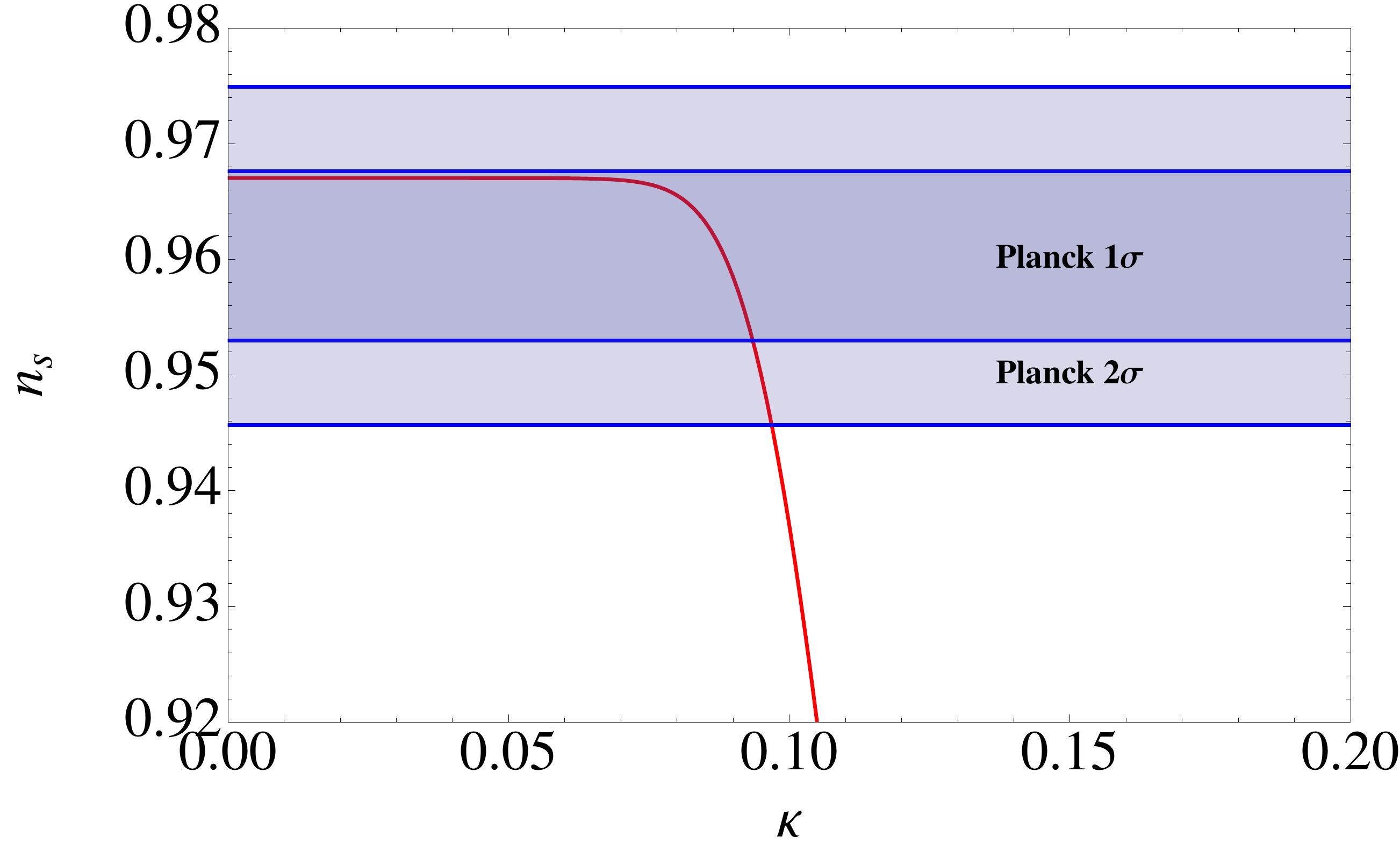}
\caption{The spectral index $n_s$ for different values of the local curvature $\kappa$. The parameters used are $\phi_0 = 0.3 \> M_{\rm pl}$, $\chi_0=10^{-3} \> M_{\rm pl}$, $\xi_\phi = 1000$, $\lambda_\phi = 0.01$, $\varepsilon = 0$ and $\Lambda_\chi = 0$.  Comparing this with Fig. \ref{fig:chi} we see that the peak in the $\beta_{\rm iso}$ curve occurs within the {\it Planck} allowed region. }
\label{fig:ns}
\end{figure}

This is indeed what we find when we study the exact numerical evolution of $n_s$ over a wide range of $\kappa$, as in Fig. \ref{fig:nswide}, as well as in the regime of weak curvature, $\kappa \ll 1$, as shown in Fig. \ref{fig:ns}. For $\kappa \lesssim 0.1$ and using $N_{\rm hc} = 60$, we find $n_s$ well within the present bounds from the {\it Planck} measurements: $n_s = 0.9603 \pm 0.0073$ \cite{PlanckInflation}. Moreover, because the regime $\kappa \lesssim 0.1$ corresponds to $T_{\cal RS} \ll 1$, the analysis of the running of the spectral index, $\alpha$, remains unchanged from \cite{KS}, and we again find $\alpha \sim -2 / N_*^2 \sim {\cal O } (10^{-4})$, easily consistent with the constraints from {\it Planck}, $\alpha = -0.0134 \pm 0.0090$ \cite{PlanckInflation}.

\begin{figure}
\includegraphics[width = 0.45\textwidth]{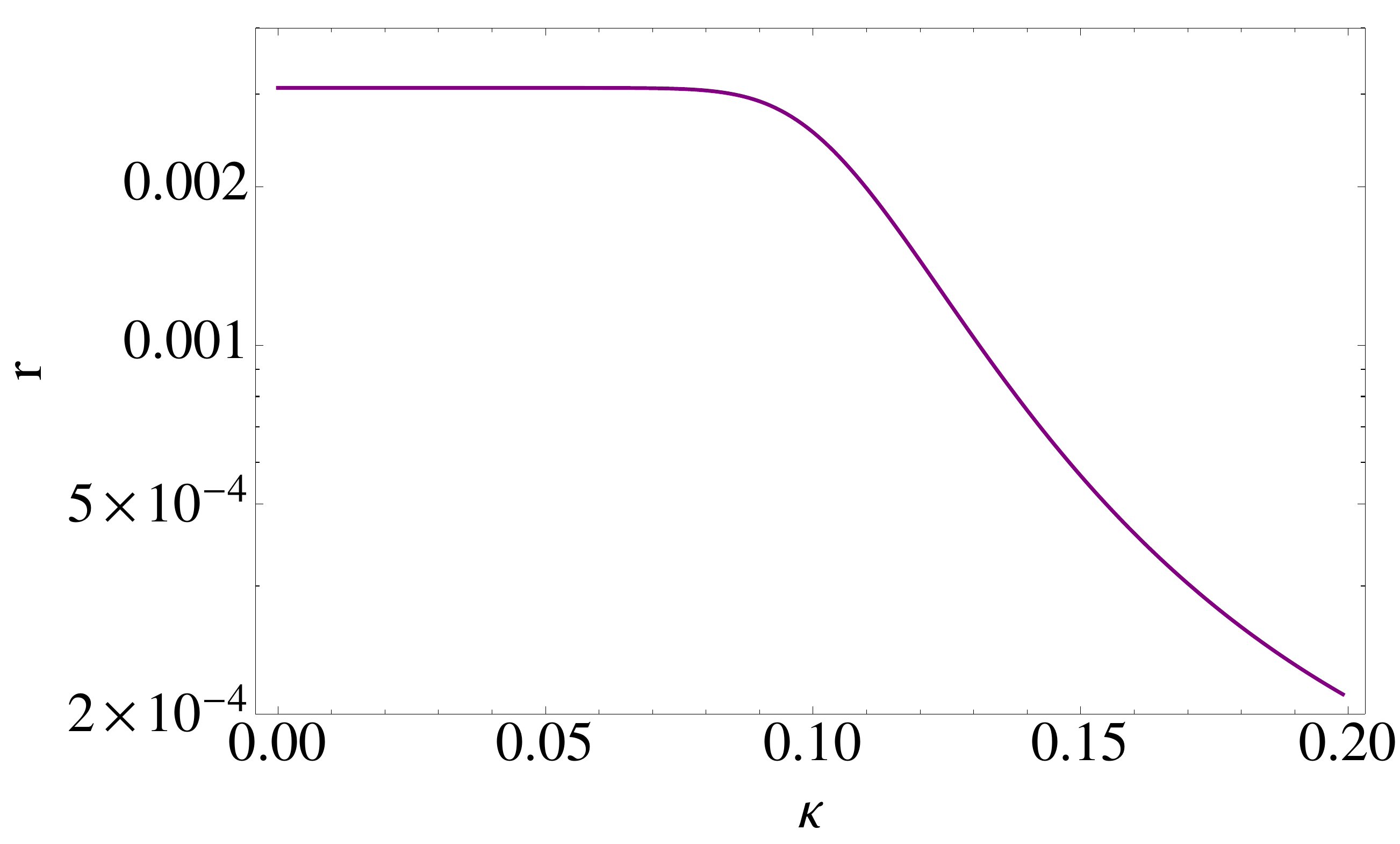}
\caption{The tensor-to-scalar ratio as a function of the local curvature parameter $\kappa$. The parameters used are $\phi_0 = 0.3 \> M_{\rm pl}$, $\chi_0=10^{-3} \> M_{\rm pl}$, $\xi_\phi = 1000$, $\lambda_\phi = 0.01$, $\varepsilon = 0$ and $\Lambda_\chi = 0$.}
\label{fig:r}
\end{figure}

\begin{figure}
\includegraphics[width = 0.45\textwidth]{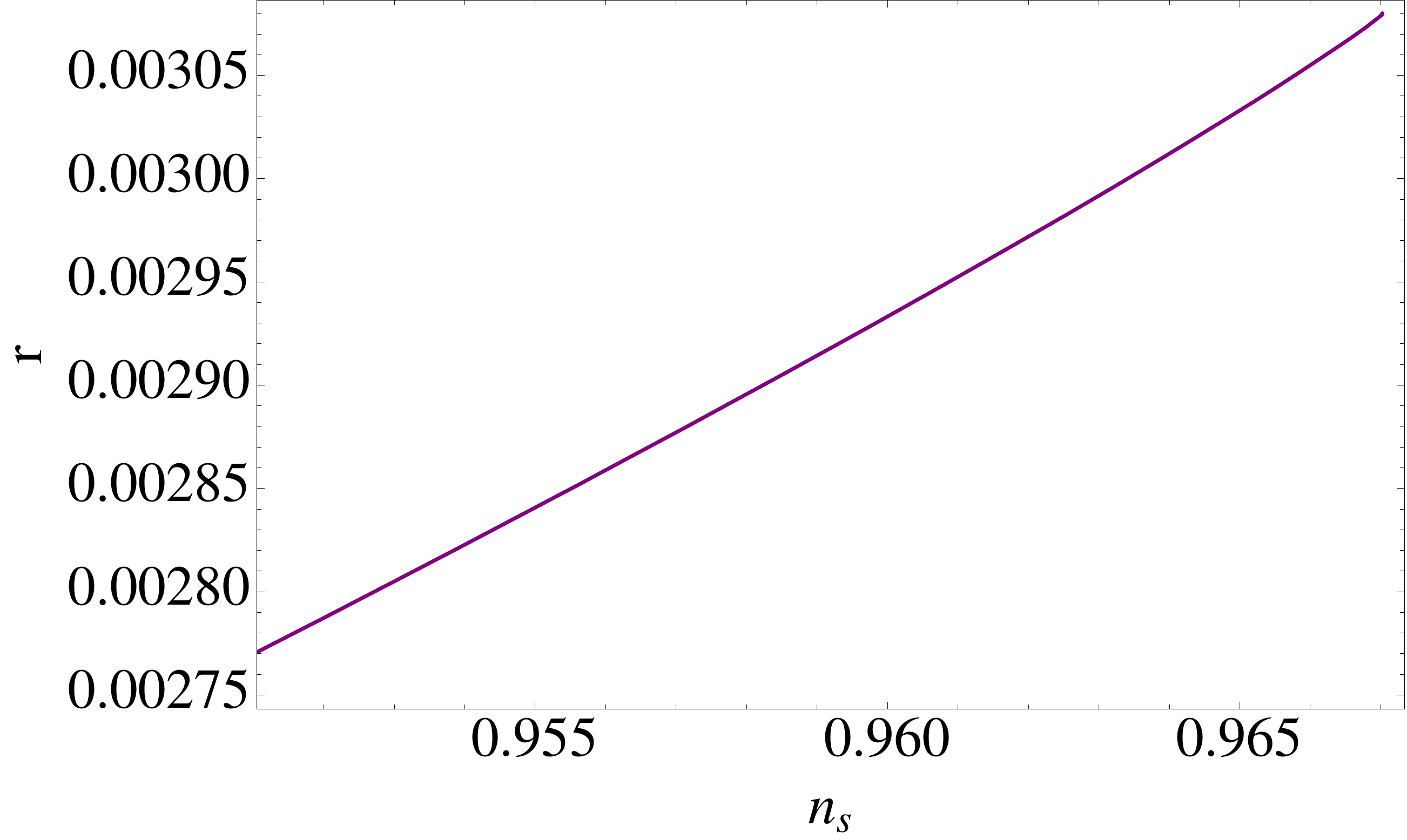}
\caption{The correlation between $r$ and $n_s$ could theoretically break the degeneracy between our models. The parameters used for this plot are $\phi_0 = 0.3 \> M_{\rm pl}$, $\chi_0=10^{-3} \> M_{\rm pl} $, $\xi_\phi = 1000$, $\lambda_\phi = 0.01$, $\varepsilon = 0$ and $\Lambda_\chi = 0$, with $0 \leq \kappa \leq 0.1$. }
\label{fig:rns}
\end{figure}

Another important observational tool for distinguishing between inflation models is the value of the tensor-to-scalar ratio, $r$. Although the current constraints are at the $10^{-1}$ level, future experiments may be able to lower the sensitivity by one or two orders of magnitude, making exact predictions potentially testable. For our models the value of $r$ is given by \cite{KS} 
\beq
r = {16 \epsilon \over 1+T_{\cal RS}^2} .
\eeq
We see that once $T_{\cal RS} \sim {\cal O} (1)$, the value of $r$ decreases, as is depicted in Fig. \ref{fig:r}.
One possible means to break the degeneracy between this family of models, apart from $\beta_{\rm iso}$, is the correlation between $r$ and $n_s$. In the limit of vanishing $T_{\cal RS}$, both $n_s$ and $r$ revert to their single-field values, though they both vary in calculable ways as $T_{\cal RS}$ grows to be ${\cal O} (1)$. See Fig. \ref{fig:rns}.

We studied the behavior of $f_{\rm NL}$ in our family of models in detail in \cite{KMS}. There we found that substantial $f_{\rm NL}$ required a large value of $T_{\cal RS}$ by the end of inflation. In this paper we have found that $T_{\cal RS}$ remains small in the regime of weak curvature, $\kappa \lesssim 0.1$. Using the methods described in detail in \cite{KMS}, we have evaluated $f_{\rm NL}$ numerically for the broad class of potentials and trajectories described above, in the limit of weak curvature ($\kappa \ll 1$), and we find $\vert f_{\rm NL} \vert \ll {\cal O} (1)$ for the entire range of parameters and initial conditions, fully consistent with the latest bounds from {\it Planck} \cite{PlanckNonGauss}.

Thus we have found that there exists a range of parameter space in which multifield dynamics remain nontrivial, producing $\beta_{\rm iso} \sim {\cal O} (0.1)$, even as the other important observable quantities remain well within the most-favored region of the latest observations from {\it Planck}. 
 

\section{Conclusions}
\label{conclusions}

Previous work has demonstrated that multifield inflation with nonminimal couplings provides close agreement with a number of spectral observables measured by the {\it Planck} collaboration \cite{KS} (see also \cite{Kallosh:2013daa}). In the limit of strong curvature of the effective potential in the Einstein frame, $\kappa \gg 1$, the single-field attractor for this class of models pins the predicted value of the spectral index, $n_s$, to within $1\sigma$ of the present best-fit observational value, while also keeping the tensor-to-scalar ratio, $r$, well below the present upper bounds. In the limit of $\kappa \gg 1$, these models also generically predict no observable running of the spectral index, and (in the absence of severe fine-tuning of initial conditions \cite{KMS}) no observable non-Gaussianity, $\vert \alpha \vert , \vert f_{\rm NL} \vert \ll 1$. In the limit of the single-field attractor, however, these models also predict no observable multifield effects, such as amplification of primordial isocurvature modes, hence $\beta_{\rm iso} \sim 0$ in the limit $\kappa \gg 1$.

In this paper, we have demonstrated that the same class of models can produce significant isocurvature modes, $\beta_{\rm iso} \sim {\cal O} (0.1)$, in the limit of weak curvature of the Einstein-frame potential, $\kappa \leq 0.1$. In that limit, these models again predict values for $n_s$, $\alpha$, $r$, and $f_{\rm NL}$ squarely within the present best-fit bounds, while also providing a plausible explanation for the observed anomaly at low multipoles in recent measurements of CMB temperature anisotropies \cite{PlanckInflation}. These models predict non-negligible isocurvature fractions across a wide range of initial field values, with a dependence of $\beta_{\rm iso}$ on couplings that admits an analytic, intuitive, geometric interpretation. Our geometric approach provides an analytically tractable method in excellent agreement with numerical simulations, which could be applied to other multifield models in which the effective potential is ``lumpy."

The mechanism for generating $\beta_{\rm iso} \sim 0.1$ that we have investigated in this paper is based on the idea that a symmetry among the fields' bare couplings $\lambda_I$, $g$, and $\xi_I$ is softly broken. Such soft breaking would result from a coupling of one of the fields (say, $\chi$) to either a CDM scalar field or to a neutrino species; some such coupling would be required in order for the primordial isocurvature perturbations to survive to the era of photon decoupling, so that the primordial perturbations could be impressed in the CMB \cite{CMBIsocurvature}. Hence whatever couplings might have enabled primordial isocurvature modes to modify the usual predictions from the simple, purely adiabatic $\Lambda$CDM model might also have generated weak but nonzero curvature in the effective potential, $\kappa \ll 1$. If the couplings $\lambda_I$, $g$, and $\xi_I$ were not subject to a (softly broken) symmetry, or if the fields' initial conditions were not such that the fields began near the top of a ridge in the potential, then the predictions from this class of models would revert to the single-field attractor results analyzed in detail in \cite{KS}.

Inflation in this class of models ends with the fields oscillating around the global minimum of the potential. Preheating in such models offers additional interesting phenomena \cite{Higgspreheating}, and further analysis is required to understand how the primordial perturbations analyzed here might be affected by preheating dynamics. In particular, preheating in multifield models --- under certain conditions --- can amplify perturbations on cosmologically interesting length scales \cite{BassettDKpreheating}. Thus the behavior of isocurvature modes during preheating \cite{isocurvpreheating} requires careful study, to confirm whether preheating effects in the family of models considered here could affect any of the predictions for observable quantities calculated in this paper. We are presently studying effects of preheating in this family of models.

Finally, expected improvements in observable constraints on the tensor-to-scalar ratio, as well as additional data on the low-$\ell$ portion of the CMB power spectrum, could further test this general class of models and perhaps distinguish among members of the class.

\subsection*{Acknowledgments}
\noindent
It is a pleasure to thank Bruce Bassett, Rhys Borchert, Xingang Chen, Alan Guth, Carter Huffman, Scott Hughes, Adrian Liu, and Edward Mazenc for helpful discussions. This work was supported in part by the U.S. Department of Energy (DOE) under Contract DE-FG02-05ER41360. KS was also supported in part by an undergraduate research fellowship from the Lord Foundation and by MIT's Undergraduate Research Opportunities Program (UROP).

\appendix

\section{Approximated Dynamical Quantities}
\label{approx}

In this appendix, we present results for dynamical quantities under our approximations that $\xi_\phi, \xi_\chi \gg 1$, $\xi_\phi \phi^2 \gg M_{\rm pl}^2$, and $\chi_0 \ll M_{\rm pl}$. 

First we expand quantities associated with field-space curvature, starting with the field-space metric, $\mathcal{G}_{I J}$, using the definition from Eq. (\ref{metric}). We arrive at the following expressions:
\begin{equation}
\begin{aligned}
& \mathcal{G}_{\phi \phi} \simeq \frac{6 M_{\rm pl}^2}{\phi ^2 } \\
& \mathcal{G}_{\phi \chi} =  \mathcal{G}_{\chi \phi} \simeq \frac{6 M_{\rm pl}^2 \xi _\chi \chi  }{\xi _{\phi} \phi ^3 } \\
& \mathcal{G}_{\chi \chi} \simeq \frac{M_{\rm pl}^2}{ \xi _{\phi } \phi ^2}.\\
\end{aligned}
\end{equation}
We also find
\begin{equation}
\begin{aligned}
& \mathcal{G}^{\phi \phi} \simeq \frac{\phi ^2 }{6 M_{\rm pl}^2} \\
& \mathcal{G}^{\phi \chi} = \mathcal{G}^{\chi \phi } \simeq -\frac{  \xi _\chi \phi \chi    }{M_{\rm pl}^2} \\
& \mathcal{G}^{\chi \chi} \simeq \frac{\xi _{\phi } \phi ^2 }{M_{\rm pl}^2}.\\
\end{aligned}
\end{equation}
Next we expand the field-space Christoffel symbols, $\Gamma^I _{ J K}$, and find
\begin{equation}
\begin{aligned}
& \Gamma ^{\phi}_{\phi \phi} \simeq -\frac{1}{\phi} \\
& \Gamma ^{\phi}_{\chi \phi} = \Gamma ^{\phi}_{\phi \chi} \simeq -\frac{\xi _\chi \chi  }{\xi _{\phi } \phi ^2 } \\
& \Gamma ^{\phi}_{\chi \chi} \simeq \frac{ \xi _\chi}{\xi _{\phi} \phi   } \\
& \Gamma ^{\chi}_{\phi \phi} \simeq \frac{\xi _\chi \chi  }{ \xi _{\phi} \phi ^2} \\
& \Gamma ^{\chi}_{\chi \phi} = \Gamma ^{\chi}_{\phi \chi} \simeq -\frac{1}{\phi} \\
& \Gamma ^{\chi}_{\chi \chi} \simeq -\frac{\xi _\chi \chi  \left(2\xi _{\phi }- \xi _\chi \right)}{\xi _{\phi }^2 \phi ^2 }.
\end{aligned}
\end{equation}
The nonzero components of the field-space Riemann curvature tensor become
\beq
\begin{split}
\mathcal{R}^{\phi}_{\phi \phi \chi} &= - \mathcal{R}^{\phi}_{\phi \chi \phi} \simeq \varepsilon (\varepsilon - 1) \frac{\chi}{\phi^3} \\
\mathcal{R}^{\phi}_{\chi \phi \chi} &= - \mathcal{R}^{\phi}_{\chi \chi \phi} \simeq - \frac{\varepsilon}{6 \xi_\phi \phi^2 } \\
\mathcal{R}^{\chi}_{\phi \chi \phi} &= - \mathcal{R}^{\chi}_{\phi \phi \chi} \simeq - \frac{\varepsilon}{\phi^2} \\
\mathcal{R}^{\chi}_{\chi \phi \chi} &= - \mathcal{R}^{\chi}_{\chi \chi \phi} \simeq \varepsilon (1 - \varepsilon) \frac{\chi}{\phi^3} .
\end{split}
\eeq

We also expand dynamical quantities, beginning with the fields' velocity:
\begin{equation}
\begin{aligned}
\dot{\sigma} \simeq \frac{\sqrt{2 \lambda _{\phi }}  M_{\rm pl}^4 }{3 \xi _{\phi }^2 \phi^2} ,
\end{aligned}
\end{equation}
and the turn rate $\omega$ in the $\phi$ and $\chi$ directions:
\begin{equation}
\begin{aligned}
& \omega^{\phi} \simeq 0\\
&  \omega^{\chi} \simeq  \frac{3 \phi^2 \left(2  M_{\rm pl} \Lambda_{\phi} \chi   -\sqrt{3 \lambda _{\phi }}  \xi _{\phi }  \dot{\chi }\right)}{2
   \sqrt{2\lambda _{\phi }}  M_{\rm pl}^3 } .
\end{aligned}
\end{equation}

\section{Covariant formalism and potential topography}

We have defined the character of the maxima and minima of the potential using the (normal) partial derivative at asymptotically large field values, where the manifold is asymptotically flat, hence the normal and covariant derivatives asymptote to the same value. By keeping the next to leading order term in the series expansion, we can test the validity of this approach for characterizing the nature of the extrema.

We take as an example the potential parameters used in Fig. \ref{fig:ev}, specifically $\xi_\phi=1000,~\xi_\chi=999.985,~\lambda_\phi=0.01,~\lambda_\chi=0.01,~g=0.01$. The ridge of the potential occurs at $\chi=0$. 

The asymptotic value of the second partial derivative is
\beq
V_{,\chi \chi} |_{\chi=0} \to  {-M_{\rm pl}^4 \Lambda_\phi \over \xi_\phi^3 \phi^2} = {-M_{\rm pl}^4 \times 1.5\cdot 10^{-5} \over \xi_\phi \phi^2}
\eeq
Let us look at the partial second derivative for $\chi=0$ and finite $\phi$:
\beqn
\nonumber
V_{,\chi \chi} |_{\chi=0} =M_{\rm pl}^4 \phi^2{ \left [ -\Lambda_\phi \xi_\phi \phi^2 + g \xi_\phi M_{\rm pl}^2 \right ]  \over \xi_\phi (M_{\rm pl}^2+\xi_\phi \phi^2)^3}
\\
\propto \left [ -0.015 ~\xi_\phi \phi^2 + 10  M_{\rm pl}^2 \right ] .
\eeqn
We see that the two terms can be comparable. In particular, the second derivative changes sign at
\beq
V_{,\chi \chi} |_{\chi=0}  =0 \Rightarrow \xi_\phi \phi_{tr}^2 \approx 667M_{\rm pl}^2
\eeq
which is a field value larger than the one we used for our calculation.  In order to get $70$ efolds of inflation, $\xi_\phi \phi^2 \sim 100M_{\rm pl}^2$, significantly smaller than the transition value. For $\phi<\phi_{tr}$ the second derivative is positive, meaning there is a transition where the local maximum becomes a local minimum. This means that if one was to take our Einstein frame potential as a phenomenological model without considering the field space metric, even at large field values, where slow roll inflation occurs, the results would be qualitatively different.

Let us now focus our attention on the covariant derivative, keeping in mind that in a curved manifold it is a much more accurate indicator of the underlying dynamics. 
\beq
{\cal D}_{\chi\chi} V = V_{,\chi\chi}-\Gamma^\phi_{\chi\chi} V_{,\phi} - \Gamma^\chi_{\chi\chi} V_{,\chi} . 
\eeq
Looking at the extra terms and keeping the lowest order terms we have $V_{,\chi} =0$ by symmetry, $V_{,\phi} \approx \lambda_\phi / ( \xi_\phi^3 \phi^3) $, and $\Gamma^\phi_{\chi\chi} = \xi_\phi (1+6\xi_\chi) \phi / C  \approx \xi_\chi / ( \xi_\phi \phi)$. 

We will now expand the covariant derivative term in $1 / \phi$ and also in $\xi_\phi$ and $\xi_\chi$. This way we will make sure that there is no transition in the behavior of the extremum for varying field values, that is to say the character of the extremum will be conserved term by term in the expansion (we only show this for the first couple of terms, but the trend is evident). We find
\beqn
\begin{split}
{\cal D}_{\chi\chi}  V &=  {- \Lambda_\phi M_{\rm pl}^4 \over \xi_\phi^3 \phi^2} \\
& + {M_{\rm pl}^6 \over \xi_\phi^3 \phi^2 (\xi_\phi \phi^2)} \left [ 2 \Lambda_\phi  -  {\lambda_\phi \varepsilon \over 6  }\left(1 - \frac{1}{6 \xi_\phi}\right)  +...\right ] \\
&+ {M_{\rm pl}^8 \over \xi_\phi^3 \phi^2 (\xi_\phi \phi^2)^2} \times \\
&\quad\quad \left [  -3\Lambda_\phi  +    {\lambda_\phi  \over 6} (1 + 2 \varepsilon )   -{\lambda_\phi \over 36 \xi_\phi} (1 + \varepsilon) +... \right ] \\
& +...
\end{split}
\label{dvexp}
\eeqn

We have written the covariant derivative using the geometrically intuitive combinations of parameters, which was done in the main text in a more general setting ($\chi \ne 0$). It is worthwhile to note that we did not write the closed form solution for ${\cal D}_{\chi\chi}  V$ (which is straightforward to calculate using the Christoffel symbols, given explicitly in \cite {KMS}), since this power series expansion is both more useful and more geometrically transparent, since it is easy to see the order at which each effect is first introduced.

We see that once we take out the $(1/ \xi_\phi^3 \phi^2)$ behavior there remains a multiple series expansion as follows
\begin {itemize}
\item Series in $(1/ \xi_\phi \phi^2)$
\item Each term of the above series is expanded in inverse powers of $\xi_\phi$.
\end{itemize}

For the example of Fig. {\ref{fig:ev} the relevant quantity that defines to lowest order in $\xi_\phi$ and $\xi_\chi$ all terms of the series is $ \Lambda_\phi = 0.015$.

By inspection of the terms, we can see that for our choice of parameters the first term defines the behavior of the covariant derivative, which is also the asymptotic value of the normal second derivative that we used to characterize the character of the extremum. In the case when $\Lambda_\phi=0$ the ellipticity term $e$ is dominant. Even if $\lambda_\phi = \varepsilon = 0$ then the dominant term comes at an even higher order and is proportional to $\lambda_\phi$.

In other words, the character of the extremum is conserved if one considers the covariant derivatives. For asymptotically large field values the two coincide, since the curvature vanishes. It is thus not only quantitatively but also qualitatively essential to use our covariant formalism for the study of these models, even at large field values where the curvature of the manifold is small.

Now that the character of the maximum is clear we can proceed to calculating all $\eta_{ss}$.
We neglect the term in ${\cal M}_{ss}$ that is proportional to ${\cal R}^I_{JKL}$, since the curvature of the field-space manifold is subdominant for $\xi_\phi \phi_0^2 \gg M^2_{\rm pl}$ and the ${\cal R}^I_{JKL}$ term is multiplied by two factors of the fields' velocity. If in addition we take $\chi = \dot \chi =0$, then ${\cal M}_{ss}$ becomes
\beq
{\cal M}_{ss} \simeq \hat{s}^{\chi} \hat{s}^\chi {\cal D}_{\chi\chi}V =  {\xi_\phi \phi^2 \over M_{\rm pl} ^2} \left ( 1 + {M_{\rm pl} ^2 \over \xi_\phi \phi^2} \right )  {\cal D}_{\chi\chi}V .
\eeq 
Using the double series expansion of Eq. (\ref{dvexp}) the entropic mass-squared becomes
\beqn
\begin{split}
{\cal M}_{ss}   V &=  {- \Lambda_\phi M_{\rm pl}^2 \over \xi_\phi^2 } \\
&+ {M_{\rm pl}^4 \over \xi_\phi^2 (\xi_\phi  \phi^2) } \left [  \Lambda_\phi  -  {\lambda_\phi \varepsilon \over 6 } \left( 1 - \frac{1}{6 \xi_\phi} \right)  +...\right ] \\
&+ {M_{\rm pl}^6 \over \xi_\phi^2 (\xi_\phi \phi^2)^2} \left [  -\Lambda_\phi  +    {\lambda_\phi  \over 6} (1 + \varepsilon)   +... \right ] +...
\end{split}
\label{mssexp}
\eeqn
To find the generalized slow roll parameter $\eta_{ss}$ we need to divide by the potential, which again can be expanded in a power series for $\chi \to 0$ as
\beq
V = M_{\rm pl}^4 {\lambda_\phi \over 4\xi_\phi^2} - M_{\rm pl}^6{\lambda_\phi \over 2 \xi_\phi^3 \phi^2} + M_{\rm pl}^8{3 \lambda_\phi \over 4 \xi_\phi^4 \phi^4} +...
\eeq
The calculation of $\eta_{ss}$ is now a straightforward exercise giving
\beqn
\begin{split}
{\eta}_{ss} & \approx {M_{\rm pl}^2 {\cal M}_{ss} \over V} =  {- 4 \Lambda_\phi  \over \lambda_\phi } \\
&\quad\quad\quad\quad + {M_{\rm pl}^2 \over \xi_\phi  \phi^2 } \left [ {- 4 \Lambda_\phi \over \lambda_\phi } -  {2 \varepsilon \over 3  }  + {\cal O} \left (  {1 \over \xi_\phi} \right ) \right ] \\
&\quad\quad\quad\quad + {M_{\rm pl}^4 \over  (\xi_\phi \phi^2)^2} \left [    {2\over 3} (1  -  \varepsilon ) + {\cal O} \left ( {1\over \xi_\phi} \right ) \right ] \\
&\quad\quad\quad\quad +{\cal O} \left ( {1\over( \xi_\phi \phi^2 )^3  } \right ) \\
& \approx - \kappa + {3\over 4N_* } \left [ {- \kappa } -  {2 \varepsilon \over 3}  \right ] + {9 \over 16 N_* ^2 } \left [    {2\over 3} (1 -  \varepsilon )  \right ] 
\end{split}
\label{etassexp}
\eeqn
where we used the slow-roll solution for $\phi$ from Eq. (\ref{phiN}), identifying it as the inflationary clock and the definition $\kappa = 4 \Lambda_\phi / \lambda_\phi$. By setting $\kappa = \varepsilon=0$ we see that even in the fully symmetric case the isocurvature mass is small but positive.  

In the limit of $\chi \to 0$ there is no turning ($\omega = 0$), and hence $T_{\cal RS} = 0$. In order to calculate $T_{SS}$ we need 
\beqn
\nonumber
\beta &=& -2\epsilon - \eta_{ss} + \eta_{\sigma\sigma} 
\\
&\simeq&  \kappa + {1 \over N_* } \left [ { 3\kappa\over 4 } + { \varepsilon \over 2}  -1 \right ] + {1 \over  N_* ^2 } \left [    {3\varepsilon \over 8} - {9\over 8}   \right ] .
\label{betaexp}
\eeqn
From Eq. (\ref{trans}), we see that $T_{\cal SS}$ depends on the integral
\beq
\int_{t_{\rm hc}}^{t} \beta H dt' = \int_{N_*}^{N_{\rm hc}} \beta dN'   .
\eeq
Plugging in the expression for $\beta$ from Eq. (\ref{betaexp})
\beq
\begin{split}
 \int_{N_*}^{N_{\rm hc}} \beta dN'   &= \kappa  (N_{\rm hc} - N_*) \\
 &\quad - c_1 \ln \left( {N_{\rm hc} \over N_*} \right) -c_2 \left (  {1\over N_*} -{1\over N_{\rm hc}}    \right )
\end{split}
\eeq
where
\beqn
c_1 &=& 1 - {3\kappa \over 4} - {\varepsilon \over 2} 
\\
c_2 &= &{9\over 8} - {3\varepsilon \over 8} .
\eeqn
Of course there is the ambiguity of stopping the integration one e-fold before the end of inflation. If one plots $\beta$ vs. $N_*$ and does a rough integration of the volume under the curve, one finds this area giving an extra contribution $\int_1 ^0 \beta dN \sim -1$. This is a change, but not a severe one. We will neglect it for now, keeping in mind that there is an ${\cal O} (1)$ multiplicative factor missing from the correct result. However since $\beta$ varies over a few orders of magnitude, we can consider this factor a small price to pay for such a simple analytical result.


\begin{thebibliography}{999}
\bibitem{HiggsMultifield} Higgs inflation should be treated as a multifield model, and hence as a member of the class under consideration here, since the Goldstone modes remain in the spectrum at high energies in renormalizable gauges \cite{WeinbergQFT,GoldstoneHE,GKS}.
\end{thebibliography}

\end{document}